\documentclass[prd,nofootinbib,preprint,superscriptaddress]{revtex4}
\pdfoutput=1
\usepackage[T1]{fontenc}
\usepackage{amsmath,amssymb}
\usepackage{epsfig}
\usepackage{subfigure}
\usepackage{float}
\usepackage{graphicx}
\usepackage{bigints}
\usepackage[usenames,dvipsnames]{color}

\usepackage{slashed}
\usepackage{multirow}
\usepackage[colorlinks,citecolor=blue]{hyperref}
\usepackage{pdfpages}
\usepackage{color}
\usepackage{comment}

\newcommand{\mhsqr}[1]{M_{H_1}^2}

\DeclareUnicodeCharacter{2212}{-}

\begin{document}
\title{Light Dirac neutrino portal dark matter with \\gauged $U(1)_{B-L}$ symmetry}

\author{Nayan Das}
\email{nayan.das@iitg.ac.in}
\affiliation{Department of Physics, Indian Institute of Technology
Guwahati, Assam 781039, India}

\author{Debasish Borah}
\email{dborah@iitg.ac.in}
\affiliation{Department of Physics, Indian Institute of Technology
Guwahati, Assam 781039, India}

\begin{abstract}
We propose a gauged $U(1)_{B-L}$ version of the light Dirac neutrino portal dark matter. The $U(1)_{B-L}$ symmetry provides a UV completion by naturally accommodating three right handed neutrinos from anomaly cancellation requirements which, in combination with the left handed neutrinos, form the sub-eV Dirac neutrinos after electroweak symmetry breaking. The particle content and the gauge charges are chosen in such a way that light neutrinos remain purely Dirac and dark matter, a gauge singlet Dirac fermion, remain stable. We consider both thermal and non-thermal production possibilities of dark matter and correlate the corresponding parameter space with the one within reach of future cosmic microwave background (CMB) experiments sensitive to enhanced relativistic degrees of freedom $\Delta N_{\rm eff}$. The interplay of dark matter, CMB, structure formation and other terrestrial constraints keep the scenario very predictive leading the $U(1)_{B-L}$ parameter space into tight corners.
\end{abstract}
\maketitle
\section{Introduction}
\label{sec:Intro}
The observations of dark matter (DM) in astrophysics and cosmology related experiments together with non-zero neutrino mass and mixing provide strong evidence for beyond standard model (BSM) physics \cite{Zyla:2020zbs, Aghanim:2018eyx}. Just like the particle nature of DM is not yet known, there are several unknowns in neutrino physics, including the origin of neutrino mass. The nature of neutrinos: Dirac or Majorana, is one of them. While there exist several BSM proposals for particle DM, the weakly interacting massive particle (WIMP) and feebly interacting massive particle (FIMP) scenarios have been studied extensively in the literature. In a typical WIMP scenario\footnote{A recent review of WIMP type scenarios can be found in \cite{Arcadi:2017kky}.}, a particle DM candidate having mass and interaction strength with standard model (SM) particles typically around the electroweak ballpark can give rise to the observed DM abundance after thermal freeze-out. On the other hand, in FIMP paradigm\footnote{A recent review of FIMP can be found in \cite{Bernal:2017kxu}.}, the DM can never enter equilibrium with the SM bath in the early universe due to its feeble interactions with the latter. Such a DM candidate, with negligible initial abundance, freezes in by virtue of decay or scattering from other particles in the bath.

In this work, we consider a scenario where the origin of DM is related to the Dirac nature of light neutrinos, known as the light Dirac neutrino portal DM \cite{Biswas:2021kio, Biswas:2022vkq} scenario\footnote{Neutrino portal DM has been studied in several earlier works where either SM light neutrino or heavy neutrinos were considered as the portal \cite{Falkowski:2009yz, Macias:2015cna, Batell:2017rol, Batell:2017cmf, Bandyopadhyay:2018qcv, Chianese:2018dsz, Blennow:2019fhy, Lamprea:2019qet, Chianese:2019epo, Bandyopadhyay:2020qpn, Hall:2019rld, Berlin:2018ztp}.}. In such a setup, light Dirac neutrinos take the role of mediating the interactions between DM and the SM bath. In \cite{Biswas:2021kio} and \cite{Biswas:2022vkq}, the DM was assumed to be of WIMP and FIMP type respectively. In addition to linking the origin of neutrino mass and nature with DM, this also offers additional discovery prospects due to right chiral part of Dirac neutrinos, contributing to the effective relativistic degrees of freedom ${\rm N_{eff}}$. Measurement related to the cosmic microwave background (CMB) puts tight constraints ${\rm N_{eff}= 2.99^{+0.34}_{-0.33}}$ at $2\sigma$ or $95\%$ CL including baryon acoustic oscillation (BAO) data \cite{Planck:2018vyg}. Similar bound also exists from big bang nucleosynthesis (BBN) $2.3 < {\rm N}_{\rm eff} <3.4$ at $95\%$ CL \cite{Cyburt:2015mya}. Both of these cosmological bounds are consistent with the SM predictions ${\rm N^{SM}_{eff}}=3.045$ \cite{Mangano:2005cc, Grohs:2015tfy,deSalas:2016ztq}\footnote{A very recent paper \cite{Cielo:2023bqp} reports $N_{\rm eff}^{\rm SM}=3.043$ by incorporating next-to-leading order correction to $e^{+}e^{-} \leftrightarrow \nu_{L}\Bar{\nu}_{L}$ interactions along with finite temperature QED corrections to the electromagnetic plasma density and effect of neutrino oscillations.} . Future CMB experiment CMB Stage IV (CMB-S4) is expected reach a much better sensitivity of $\Delta {\rm N}_{\rm eff}={\rm N}_{\rm eff}-{\rm N}^{\rm SM}_{\rm eff}
= 0.06$ \cite{Abazajian:2019eic}, taking it closer to the SM prediction. Thus, Dirac neutrino scenarios can be probed in future CMB experiments if right handed neutrinos (RHN) can be sufficiently produced in the early universe via thermal or non-thermal processes. Light Dirac neutrino models often lead to enhanced $\Delta {\rm N}_{\rm eff}$, some recent works on which can be found in \cite{Abazajian:2019oqj, FileviezPerez:2019cyn, Nanda:2019nqy, Han:2020oet, Luo:2020sho, Borah:2020boy, Adshead:2020ekg, Luo:2020fdt, Mahanta:2021plx, Du:2021idh, Biswas:2021kio, Borah:2022obi, Borah:2022qln, Li:2022yna, Biswas:2022fga, Adshead:2022ovo, Borah:2023dhk, Borah:2022enh, Das:2023oph}. In earlier works on light Dirac neutrino portal DM, discrete symmetry like $Z_4$ was considered to get the desired couplings, mass terms as well as the stability of DM. Here, we consider a UV completion with a gauged $B-L$ framework. The right handed neutrinos are naturally part of the model providing the minimal anomaly-free setup. While RHNs are thermally produced due to gauged $B-L$ interactions, DM can be of WIMP or FIMP type depending upon Dirac neutrino portal couplings. The model not only gives rise to the desired DM phenomenology with observable $\Delta {\rm N}_{\rm eff}$, but also leads to new constraints in the gauged $B-L$ parameter space not obtained previously.

This paper is organised as follows. In section \ref{sec:model}, we briefly discuss the minimal gauged $U(1)_{B-L}$ model of light Dirac neutrino portal dark matter followed by separate discussions of FIMP and WIMP type DM in section \ref{sec:fimp} and \ref{sec:wimp} respectively. We finally conclude in section \ref{sec:conclude}.

\section{The Model}
\label{sec:model}
Gauged $B-L$ extension of the SM \cite{Davidson:1978pm, Mohapatra:1980qe, Marshak:1979fm, Masiero:1982fi, Mohapatra:1982xz, Buchmuller:1991ce} has been a popular BSM framework studied in the context of neutrino mass among others. The three right handed neutrinos $\nu_R$ having $B-L$ charge -1 each not only keep the model anomaly free\footnote{See appendix \ref{appen_anomaly} for other solutions to anomaly cancellation conditions.} but also lead to massive Dirac neutrinos in combination with $\nu_L$ after electroweak symmetry breaking. A singlet fermion $\psi$ is considered to be the DM candidate while two singlet scalars $\phi_{1}$ and $\phi_{2}$ with non-zero $B-L$ charges help in realising light Dirac neutrino portal DM and spontaneous $B-L$ symmetry breaking respectively. The relevant particle content is shown in table \ref{tab1}.

\begin{table}[h]
        \centering
	\begin{tabular}{|c|cc|cccc|}
		\hline
		&$L$&$\Phi$&$\nu_R$&$\psi$&$\phi_1$&$\phi_2$\\
		\hline
		$SU(2)$&2&2&1&1&1&1\\
		$U(1)_Y$&$-\frac{1}{2}$&$\frac{1}{2}$&0&0& 0 &0\\
   	$U(1)_{B-L}$&$-1$&$0$&-1&0&1&3\\

		\hline
	\end{tabular}
	\caption{Relevant particle content of the model with respective quantum numbers under the symmetry group.}\label{tab1}
\end{table}

The scalar Lagrangian for the model can be written as
\begin{equation}
    \mathcal{L_{\text{s}}} = (D^{\mu}\Phi)^{\dagger}D_{\mu}\Phi + (D^{\mu}\phi_{1})^{\dagger}D_{\mu}\phi_{1} + (D^{\mu}\phi_{2})^{\dagger}D_{\mu}\phi_{2} - V(\Phi, \phi_{1}, \phi_{2}),
\end{equation}
where
\begin{equation}
D_{\mu}\Phi = \left(\partial_{\mu} - i \frac{g}{2} \tau_{a} W^{a}_{\mu} - i \frac{g'}{2} B_{\mu}\right)\Phi, \, D_{\mu}\phi_{1} = \left(\partial_{\mu} - i g_{BL} B'_{\mu}\right)\phi_{1}, \, 
D_{\mu}\phi_{2} = \left(\partial_{\mu} - i 3 g_{BL} B'_{\mu}\right)\phi_{2},
\end{equation}
denote the respective covariant derivatives and
\begin{eqnarray}
    V(\Phi, \phi_{1}, \phi_{2}) &=& -\mu^2(\Phi^\dagger \Phi) + \mu_{1}^2(\phi_{1}^\dagger \phi_{1}) - \mu_{2}^2(\phi_{2}^\dagger \phi_{2}) + \lambda(\Phi^\dagger \Phi)^2 + \lambda_{1}(\phi^\dagger_{1} \phi_{1})^2 + \lambda_{2}(\phi^\dagger_{2} \phi_{2})^2  \nonumber \\
    &+& \lambda_{H\phi_{1}}(\Phi^\dagger \Phi)(\phi_{1}^\dagger \phi_{1}) + \lambda_{H\phi_{2}}(\Phi^\dagger \Phi)(\phi_{2}^\dagger \phi_{2}) + \lambda_{\phi_{1}\phi_{2}}(\phi_{1}^\dagger \phi_{1})(\phi_{2}^\dagger \phi_{2 }) \nonumber \\
    & + & {(\lambda'_{\phi_{1}\phi_{2}} \phi^3_1 \phi^\dagger_2 + {\rm h.c.})}
\end{eqnarray}
denotes the scalar potential. The relevant part of the fermion Lagrangian is
\begin{equation}
-\mathcal{L_{\text{Y}}} \supset Y_\nu \overline{L} \tilde{\Phi} \nu_R + y_{\phi_{1}} \overline{\psi} \phi_1 \nu_R +  m_\psi \overline{\psi} \psi.
\end{equation}

The singlet scalar field $\phi_2$ acquires a non-zero vacuum expectation value (VEV) denoted by $\langle \phi_2 \rangle =v_2$, leading to the spontaneous breaking of $U(1)_{B-L}$ gauge symmetry. The singlet field $\phi_1$ does not acquire any VEV and remains heavier than DM $\psi$, ensuring latter's stability. We also consider the Higgs portal coupling $\lambda_{H\phi_{2}}$ to be negligible for simplicity.


Considering the the free parameters in the model to be $\mu_{1}, \mu_{2},\lambda_{1}, \lambda_{2}, \lambda_{H\phi_{1}}, {\lambda'_{\phi_{1}\phi_{2}}}, g_{BL}, y_{\phi_{1}}, m_{\psi}$, the physical masses of $\phi_{1}, \phi_{2}$ and $Z'$ (the $U(1)_{B-L}$ gauge boson) can be written as
\begin{eqnarray}
    m_{\phi_{1}}^2 &=& \mu_{1}^2 + \frac{1}{2} \lambda_{H\phi_{1}} v^2, \\
    m_{\phi_{2}}^2 &=& 2 \lambda_{2} v_{2}^2, \\
    m_{Z'}^2 &=& 9 g_{B-L}^2 v_{2}^2,
\end{eqnarray}
where $v \left(= \sqrt{\frac{\mu^2}{\lambda}}\right)$ and $v_{2} \left(= \sqrt{\frac{\mu_{2}^2}{\lambda_{2}}}\right)$ are the VEVs of the neutral component of the SM Higgs $\Phi$ and $\phi_{2}$ respectively. As a result, the parameters $\mu_{1}, \mu_{2}$ and $\lambda_{2}$ can be traded for the parameters $m_{\phi_{1}}, m_{\phi_{2}}$ and $m_{Z'}$. This leads to the free parameters as
$m_{\phi_{1}}, m_{\phi_{2}}, m_{Z'}, \lambda_{1}, \lambda_{H\phi_{1}}, {\lambda'_{\phi_{1}\phi_{2}}} , g_{BL}, y_{\phi_{1}}, m_{\psi}$, out of which the relevant parameters for the phenomenology to be discussed are \footnote{ The scalar $\phi_{2}$ is not expected to play any role in dark matter analysis. First of all, DM does not couple directly to $\phi_2$. Also, $\phi_2$ is heavier than the rest of the particles and hence does not appear in final states.}
$m_{\phi_{1}},  m_{Z'}, m_{\psi}, \lambda_{H\phi_{1}}, g_{BL}, y_{\phi_{1}}$.

While we have a gauged $B-L$ symmetry, DM is neutral under $U(1)_{B-L}$ and its relic depends on the Dirac neutrino portal couplings in the spirit of light Dirac neutrino portal DM \cite{Biswas:2021kio, Biswas:2022vkq}.
Now, depending on the value of this Dirac neutrino portal Yukawa coupling $y_{\phi_{1}}$, the DM analysis can be broadly divided into two categories namely, (a) FIMP ($y_{\phi_{1}}<10^{-7}$) and (b) WIMP ($y_{\phi_{1}}>10^{-7}$). In the first case, due to small Yukawa coupling, the DM $\psi$ is produced non-thermally and dominantly from $\phi_1$ decay. In the second case, DM can attain equilibrium in the early universe due to sizeable interactions. We now discuss these two broad cases one by one.

\section{FIMP type dark matter}
\label{sec:fimp}
In this scenario, the Yukawa coupling among $\nu_R$, DM and singlet scalar is small $y_{\phi_{1}}<10^{-7}$ keeping the associated processes out-of-equilibrium throughout. While DM never thermalises, the RHNs can thermalise with the SM bath by virtue of $U(1)_{B-L}$ gauge interactions. Similar to the active neutrinos, the RHNs also  contributes to the effective relativistic degrees of freedom defined as
\begin{eqnarray}
 N_{\rm eff} \equiv \frac{8}{7} \left( \frac{11}{4} \right)^{4/3} \left( \frac{\rho_{\rm rad} -\rho_{\gamma}}{\rho_{\gamma}} \right),     
\end{eqnarray}
where $\rho_{\rm rad}$ is the net radiation content of the universe \footnote{Due to the presence of three relativistic RHNs, change in $N_{\rm eff}$ can be written as $\Delta N_{\rm eff} = 3 \left.{\frac{\rho_{\nu_{R}}}{\rho_{\nu_{L}}}}\right|_{\rm CMB} = 3 \left.{\frac{\rho_{\nu_{R}}}{\rho_{\nu_{L}}}}\right|_{\rm 10 MeV} = 3 \left.{\left(\frac{T_{\nu_{R}}}{T_{\nu_{L}}}\right)^{4}}\right|_{\rm 10 MeV}$. Here $\rho_{\nu_{L}}$ denotes energy density of one species of SM neutrinos. For non-thermal RHNs, $\rho_{\nu_{R}}$ should be calculated by solving appropriate Boltzmann equation whereas for thermal RHNs, $\rho_{\nu_{R}}$ can be written in terms of its decoupling temperature.}. As mentioned earlier, the SM prediction is $N^{\rm SM}_{\rm eff}=3.045$ \cite{Mangano:2005cc, Grohs:2015tfy, deSalas:2016ztq}. In our model for thermal right handed neutrinos ($\nu_{R}$), $\Delta N_{\rm eff}$ can be estimated by finding the decoupling temperature $T_{\nu_R}$ of $\nu_R$ using 
\begin{eqnarray}
\Gamma (T_{\nu_R}) = \mathcal{H} (T_{\nu_R})
\label{eq7}
\end{eqnarray}
where $\Gamma(T)$ is the interaction rate and $\mathcal{H}(T)$ is the expansion rate of the universe.

\begin{figure}
    \centering
    \includegraphics[height=8 cm, width = 10 cm]{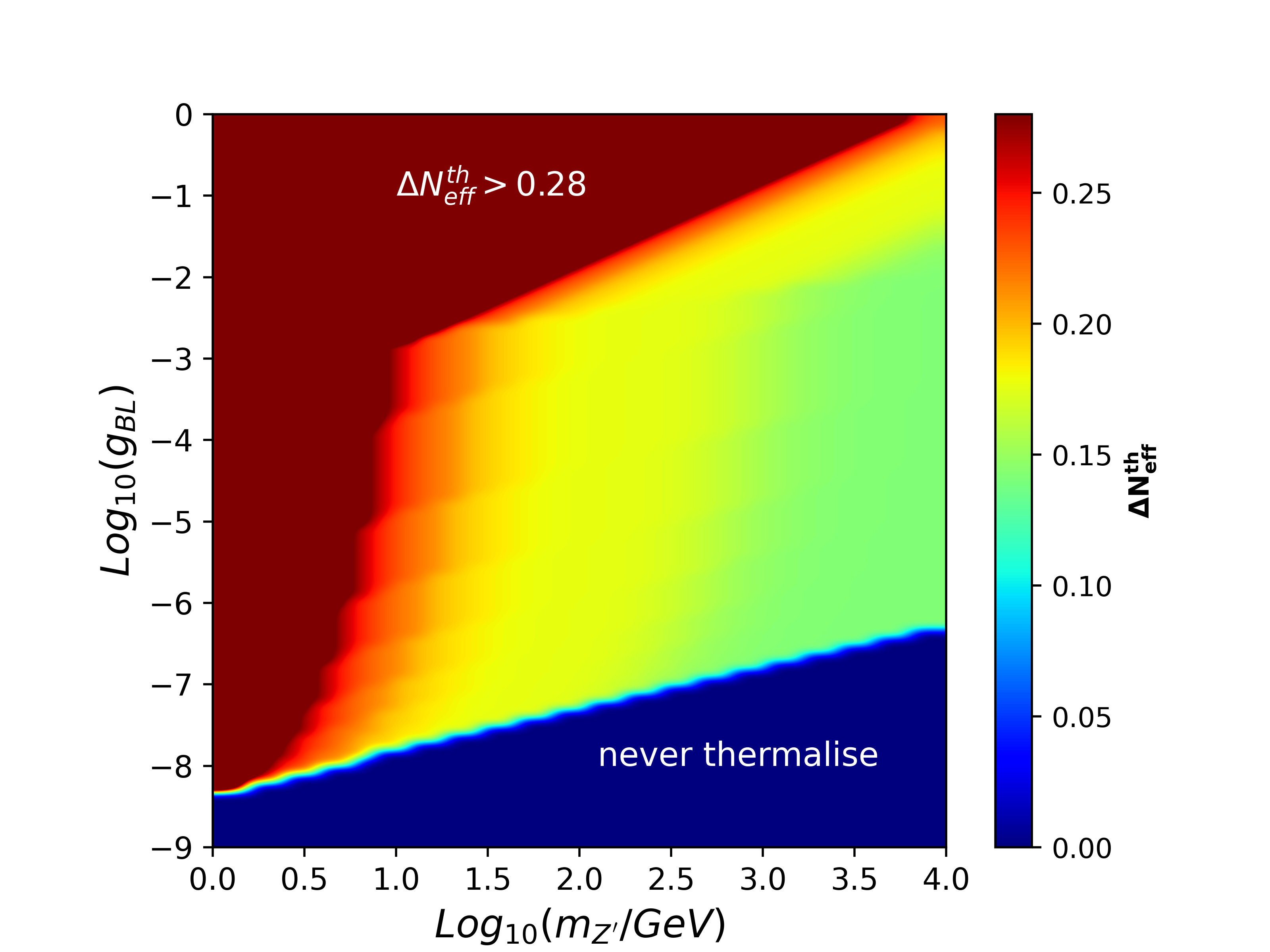}
    \caption{$\rm \Delta N^{\rm th}_{\rm eff}$ as a function of $m_{Z'}$ and $g_{BL}$.}
    \label{fig:thermal_Neff}
\end{figure}

Let us consider the themalisation of $\nu_{R}$ with the SM bath via the $Z'$ portal interactions. Sufficiently large $Z'$ portal interactions can lead to thermalisation of $\nu_{R}$ via the s-channel process $f\Bar{f} \leftrightarrow \nu_{R}\Bar{\nu}_{R}$ (here $f$ denotes SM fermions). As a result, $\nu_{R}$ enters the thermal bath and then, after a certain period, it decouples from the bath. This generates the thermal contribution of $\nu_{R}$ to $\Delta N_{\rm eff}$ \cite{Abazajian:2019oqj}. Fig. \ref{fig:thermal_Neff} shows the thermal $\Delta N_{\rm eff}$ as a function of coupling $g_{BL}$ and gauge boson mass $m_{Z'}$. Thus for $g_{BL} \gtrsim 10^{-8}$, $\Delta N_{\rm eff}$ receives a thermal contribution from light Dirac neutrinos which we denote as $\Delta N^{\rm th}_{\rm eff}$. The region corresponding to large $Z'$ portal interactions, labelled as $\Delta N_{\rm eff}> 0.28$ is disfavoured by Planck 2018 limits at $2\sigma$ CL.

On the other hand, due to the tiny dark sector Yukawa coupling $y_{\phi_1}$, it is also possible to get a non-thermal contribution to $\Delta N_{\rm eff}$ \cite{Luo:2020fdt, Biswas:2022vkq}, which we denote by $\Delta N_{\rm eff}^{\rm non-th}$. In addition to the Yukawa coupling $y_{\phi_1}$ involved in $\phi_1-\nu_R-\psi$ coupling, the $U(1)_{B-L}$ portal coupling and scalar portal coupling $\lambda_{H\phi_{1}}$ can also play a crucial role in deciding the strength of this non-thermal contribution. While the total contribution to $\Delta N_{\rm eff}$, in general, is a combination of thermal contribution and non-thermal contribution, the latter can occur only if the decoupling of RHNs precedes the non-thermal or freeze-in production.

To study this non-thermal contribution $\Delta N_{\rm eff}^{\rm non-th}$ in details, let us first consider the situation where $\phi_{1}$ remains in thermal equilibrium initially and then freezes out from the bath. The frozen out $\phi_{1}$ later decays into $\psi$ and $\nu_{R}$, leading to non-thermal production of both dark matter and $\nu_R$. The non-thermal contribution to $\Delta N_{\rm eff}$ from $\nu_{R}$ can be calculated from the total energy density of frozen-in Dirac neutrinos.

The Boltzmann equation for comoving number density of $\phi_{1}$ can be written as
\begin{equation}
    \frac{dY_{\phi_{1}}}{dx} =  \frac{\beta s}{\mathcal{H} x} \left(-\langle \sigma v\rangle _{_{\phi_{1} \phi^{\dagger{}}_{1} \to X \bar{X}}}  \left((Y_{\phi_{1}})^2 -  (Y_{\phi_{1}}^{\rm eq})^2 \right) - \frac{\Gamma_{\phi_{1}}}{s} \frac{K_1(x)}{K_2(x)} Y_{\phi_{1}}\right),
    \label{phi1_freeze_out}
\end{equation}
where $K_{1}$ and $K_{2}$ are the modified Bessel functions of first and second kind respectively. The comoving abundance of species $i$ is defined as $Y_i = n_i/s$ with $n_i, s$ being number density of species $i$ and entropy density of the universe respectively. $\mathcal{H}$ denotes the Hubble parameter. The variable $x$ is defined as $x=m_{\phi_1}/T$ and $\beta=1 + \frac{T}{3 g^{s}_*} \frac{d g^{s}_{*}}{dT}$ with $g^{s}_*$ being the relativistic entropy degrees of freedom. Here $\langle{\sigma v}\rangle _{_{\phi_{1} \phi_{1}^{\dagger{}} \to X \bar{X}}}$ is the thermally averaged annihilation cross-section of $\phi_1$ into the all allowed final state particles and $\Gamma_{\phi_{1}}$ denotes the decay rate of $\phi_{1}$ into $\nu_{R}$ and $\psi$. The annihilation cross-section of $\phi_1$ depends upon Higgs portal coupling $\lambda_{H\phi_{1}}$ as well as $U(1)_{B-L}$ portal coupling. Depending upon the strength of these individual couplings, the freeze-out abundance of $\phi_1$ is either determined by scalar portal or gauge portal interactions. The freeze-in abundance of $\psi$ and $\nu_{R}$ (from the decay of frozen out $\phi_{1}$) can be obtained by solving the relevant Boltzmann equations given by
 \begin{equation}
     \frac{dY_{\psi}}{dx} = \frac{ \beta }{x \mathcal{H} } \Gamma_{\phi_{1}} \frac{K_1(x)}{K_2(x)} Y_{\phi_{1}},
     \label{psi_freezed_out}
\end{equation}
\begin{equation}
       \frac{d \widetilde{Y}}{dx} = \frac{\beta}{\mathcal{H} s^{1/3} x} \langle E\Gamma \rangle_{\phi_{1}}  Y_{\phi_{1}},
       \label{nur_freezed_out}
\end{equation}
with 
\begin{eqnarray}
    \Gamma_{\phi_{1}} = \frac{1}{32 \pi} m_{\phi_{1}} y^2_{\phi_{1}} \left(1-\frac{m^2_{\psi}}{m^2_{\phi_{1}}}\right) ^2; \hspace{1 cm} \langle E\Gamma \rangle_{\phi_{1}} = \frac{1}{64 \pi} m^2_{\phi_{1}} y^2_{\phi_{1}} \left(1-\frac{m^2_{\psi}}{m^2_{\phi_{1}}}\right) ^3. \nonumber
\end{eqnarray}
The term $\Tilde{Y} = \frac{\rho_{\nu_{R}}}{s}$ represents the comoving energy density of RHNs. The preferred choice of energy density instead of number density of $\nu_R$ is based on the fact that the calculation of $\Delta N_{\rm eff}$ requires comoving energy density of $\nu_{R}$.

\begin{figure}[h!]
\includegraphics[height=5.5cm,width=7.5cm,angle=0]{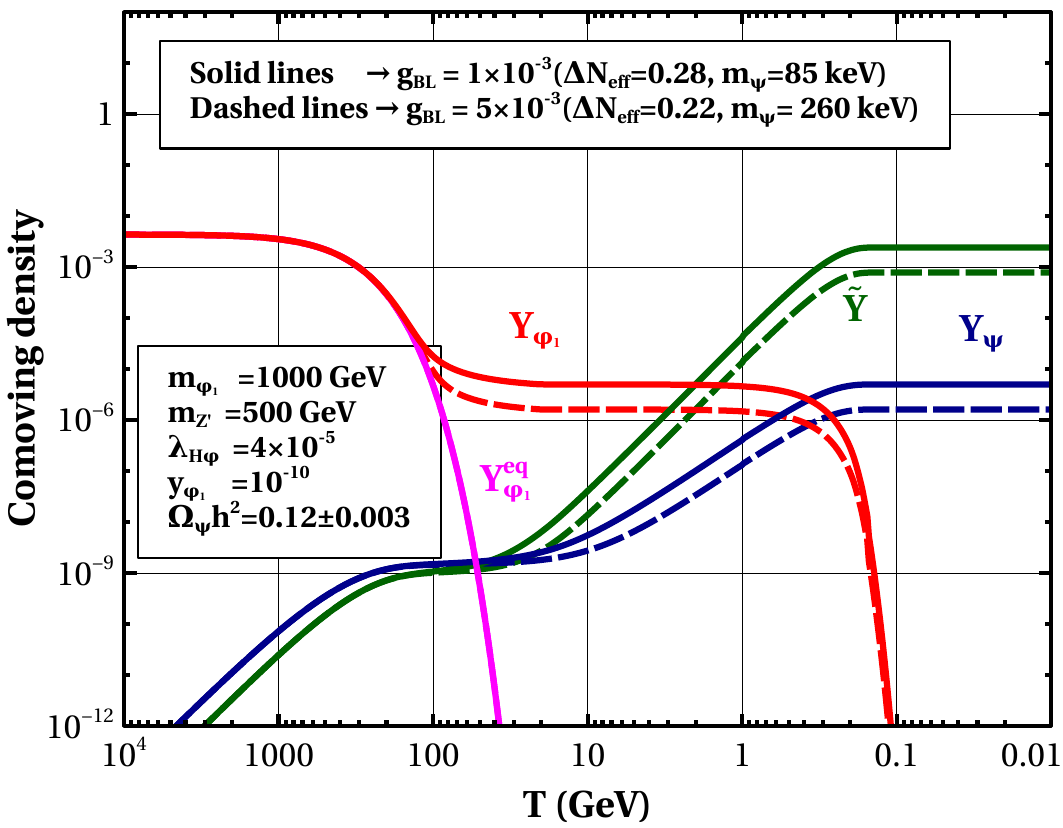}
\includegraphics[height=5.5cm,width=7.5cm,angle=0]{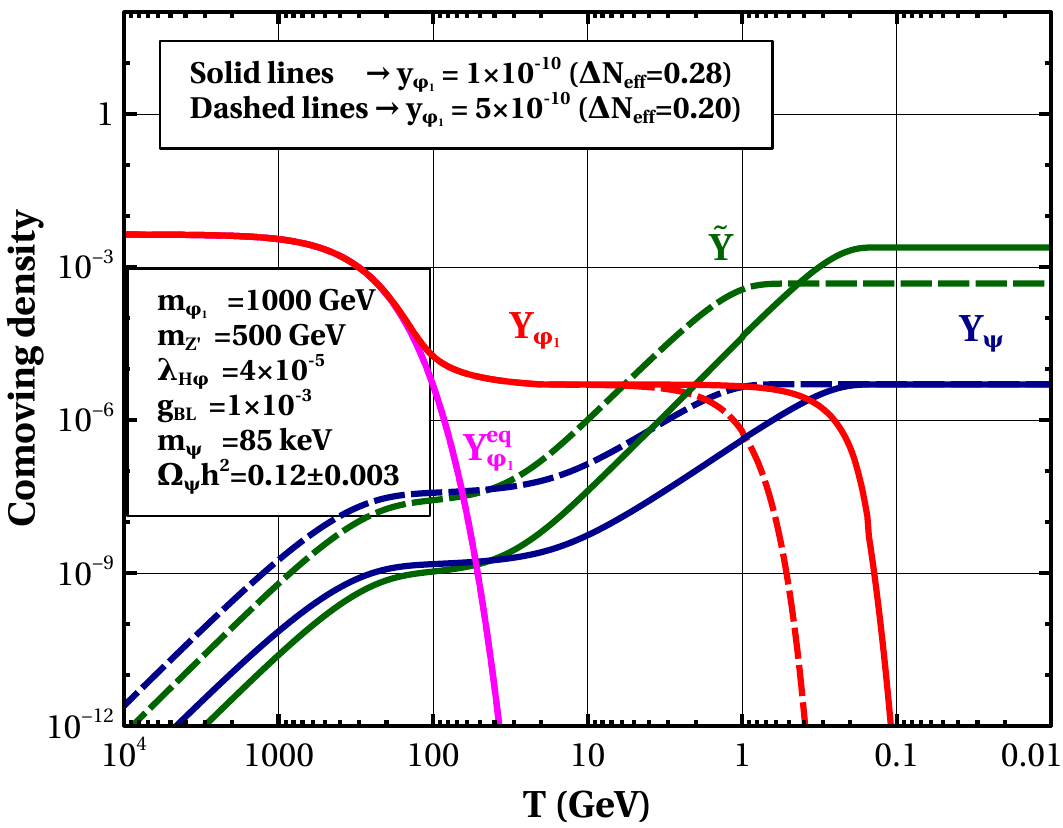}
\caption{Evolution of comoving number density of $\phi_{1}$, $\psi$ and comoving energy density of $\nu_{R}$. The left (right) panel shows the variation with respect to $g_{BL}$ ($y_{\phi_{1}}$).}
\label{fig:FIMP_comparison}
\end{figure}

The non-thermal contribution to $\Delta N_{\rm eff}$ namely, $\Delta N^{\rm non-th}_{\rm eff}$ depends on the parameters $m_{\phi_{1}},  m_{Z'},$ $\lambda_{H\phi_{1}}, g_{BL}$ and $y_{\phi_{1}}$. While the parameters $ m_{\phi_{1}}, m_{Z'}, \lambda_{H\phi_{1}}$ and $g_{BL}$ determine the freeze-out abundance of $\phi_{1}$, the parameters that determine the decay width of $\phi_{1}$ are $m_{\phi_{1}}, y_{\phi_{1}}$. Fig. \ref{fig:FIMP_comparison} shows the evolution of comoving densities for chosen benchmark points clearly indicating the roles of $g_{BL}$ and $y_{\phi_{1}}$. While we only show the evolution of non-thermal contribution to $\Delta N_{\rm eff}$ in this figure, the total $\Delta N_{\rm eff}$ for the chosen benchmark values of parameters include both thermal and non-thermal contributions. In the left panel plot of Fig. \ref{fig:FIMP_comparison}, the solid line for $\widetilde{Y}$ corresponds to an asymptotic non-thermal contribution $\Delta N^{\rm non-th}_{\rm eff}=0.10$ while the same choice of parameters generates a thermal contribution $\Delta N^{\rm th}_{\rm eff}=0.18$. Therefore, the total contribution of $\Delta N_{\rm eff}=0.28$, as indicated in the legends for the solid line. For the dashed line, corresponding to a larger value of $g_{BL}$, the thermal contribution also comes out to be slightly larger $\Delta N^{\rm th}_{\rm eff}=0.19$, as expected. However, a significant decrease in non-thermal contribution is observed for this benchmark point leading to $\Delta N^{\rm non-th}_{\rm eff}=0.03$. Thus, for the dashed line in left panel plot of Fig. \ref{fig:FIMP_comparison}, we have total $\Delta N_{\rm eff}=0.22$. For the right panel plot of the same figure, the thermal contribution remains same $\Delta N^{\rm th}_{\rm eff}=0.18$ as $g_{BL}$ and $m_{Z'}$ are kept fixed and variation in Yukawa coupling $y_{\phi_1}$ in the non-thermal ballpark does not alter thermal contribution to $\Delta N_{\rm eff}$. The dashed line, with a larger $y_{\phi_1}$, have a smaller $\Delta N^{\rm non-th}_{\rm eff}=0.02$ and vice versa. As a result, for the dashed line in right panel plot, the total contribution is $\Delta N_{\rm eff}=0.20$. In both the left and right panel plots, the values of $m_{\psi}$ are taken in such a way that it gives correct FIMP type dark matter relic. Also, we have seen that the value of $m_{\psi}$ does not affect $\Delta N_{\rm eff}$ as long as $m_{\psi} \ll m_{\phi_{1}}$. 

\begin{figure}[h!]
\includegraphics[scale=0.35]{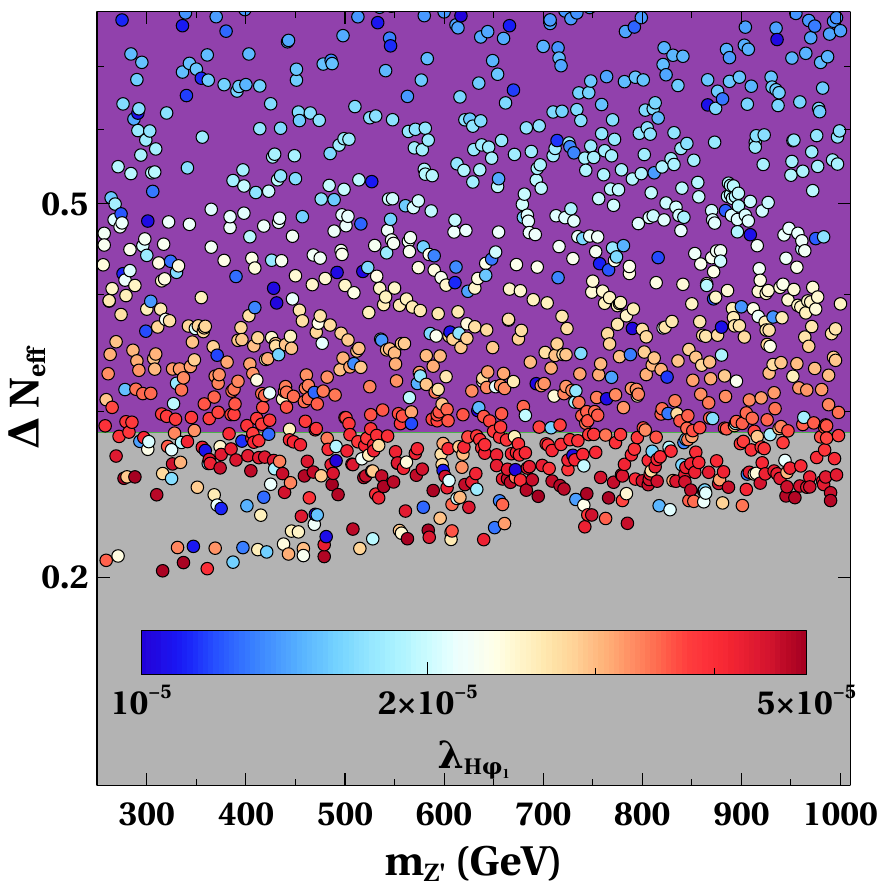}
\includegraphics[scale=0.35]{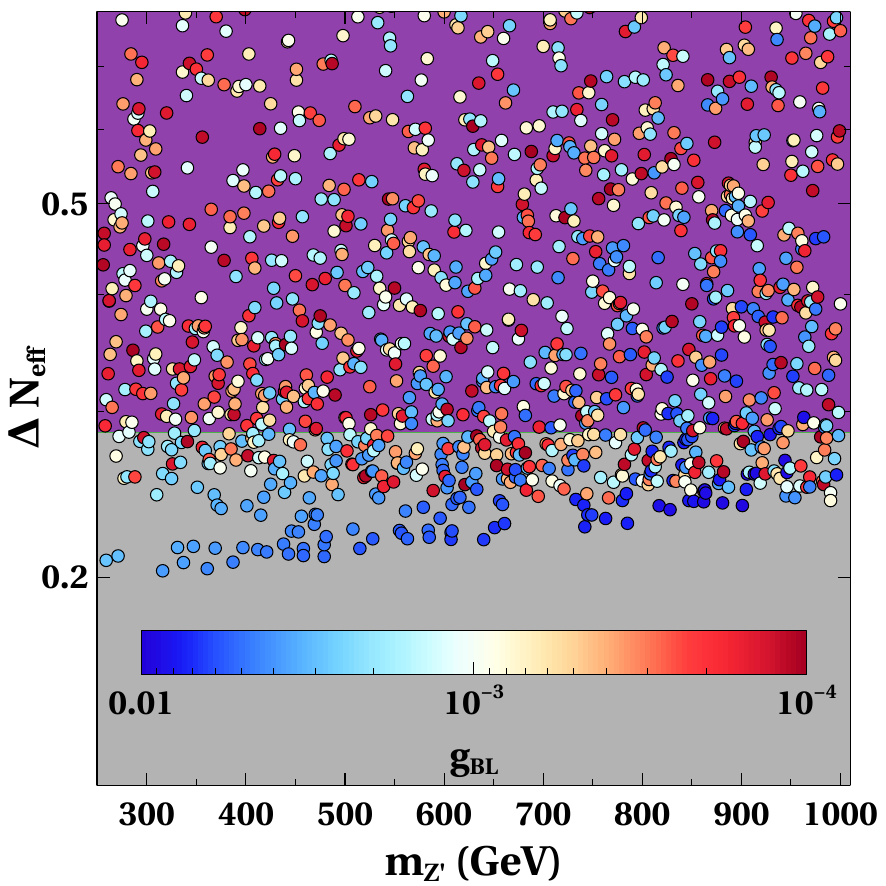}
\includegraphics[scale=0.35]{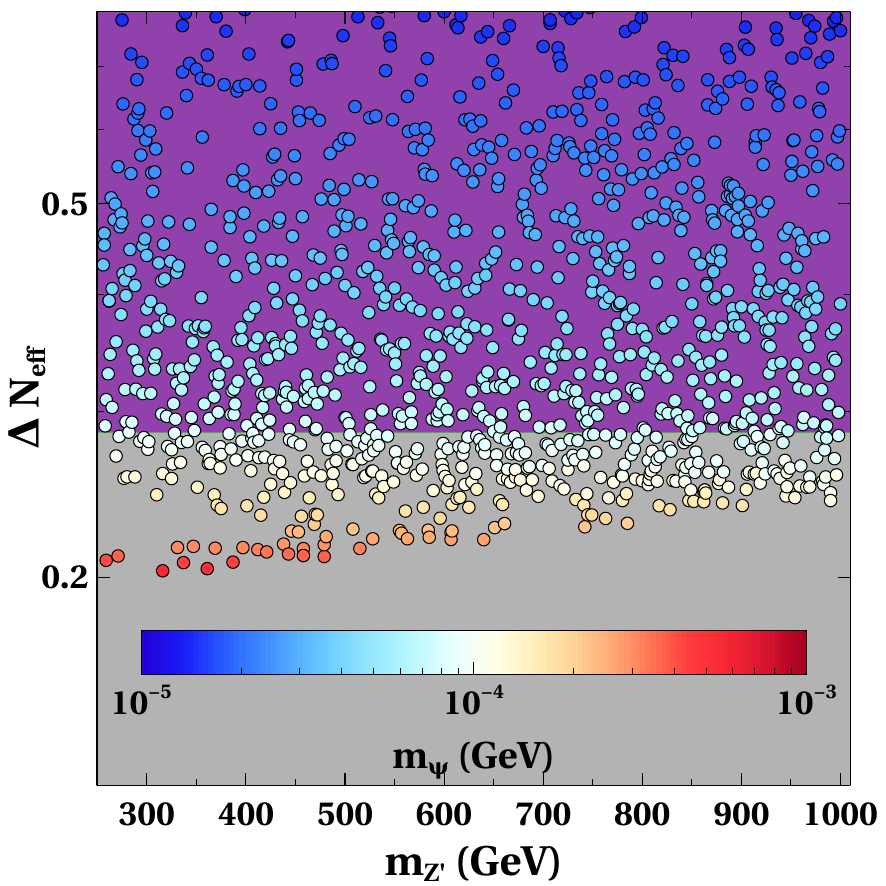}
\caption{Scan plot showing the total $\Delta N_{\rm eff}$ versus $m_{Z'}$ for different $\lambda_{H\phi_{1}}$, $g_{BL}$ and $m_{\psi}$. The magenta coloured region is excluded by Planck 2018 bounds at $2\sigma$ CL. The region in grey color is within the reach of future experiments.}
\label{fig:SY_scan}
\end{figure}

After studying the evolution of comoving densities for different benchmark points, we perform a numerical scan over key model parameters to find out the parameter space consistent with $\Delta N_{\rm eff}$ and DM properties. We have kept $m_{\phi_1}$ fixed at $1000$ GeV and $y_{\phi_{1}}$ at $10^{-10}$. The rest of the parameters are varied in the following range:
\begin{eqnarray}
    250 \, \text{GeV} &< m_{Z'} &< 1000 \, \text{GeV} \\ \nonumber
    10^{-5} &< \lambda_{H\phi_{1}} &<5\times10^{-5} \\ \nonumber
    10^{-4} &< g_{BL} &< 10^{-2} \\ \nonumber
    10^{-5} \, \text{GeV} &< m_{\psi} &< 10^{-3} \, \text{GeV}. \nonumber
\end{eqnarray}
The resulting parameter space in terms of $\Delta N_{\rm eff}$ and $m_{Z'}$ is shown in Fig. \ref{fig:SY_scan}. All the points shown in this figure satisfy the requirements of correct DM relic abundances. The magenta shaded region denotes the region excluded by Planck 2018 bounds at $2\sigma$ CL while the grey shaded region remains within the reach of future
experiments like CMB-S4. The colour codes in the left, middle and right panel plots of Fig. \ref{fig:SY_scan} show the variation in $\lambda_{H \phi_1}, g_{BL}, m_{\psi}$ respectively. As the left panel plot shows,  decrease in $\lambda_{H\phi_1}$, while keeping $m_{\phi_1}$ constant, increases $\Delta { N_{\rm eff}}$. A smaller value of Higgs portal coupling $\lambda_{H\phi_1}$ leads to a larger freeze-out abundance of $\phi_1$ followed by enhanced production of $\nu_R$ from $\phi_1$ decay. Since the same decay is also responsible for freeze-in production of DM, we require smaller DM masses in order to keep its relic abundance within Planck limits, as seen from the right panel plot of Fig. \ref{fig:SY_scan}. As the middle panel plot shows, small values of $\Delta N_{\rm eff}$ typically correspond to smaller $g_{BL}$ as the corresponding thermal contribution $\Delta N^{\rm th}_{\rm eff}$ decreases.

While we have incorporated the constraints from cosmological observations on $\Delta N_{\rm eff}$ and DM relic abundance, there can be strong constraints on light dark matter from astrophysical structure formation. Such bounds can be imposed on a particular DM scenario by calculating the free-streaming length (FSL) of DM. While hot DM is already ruled out, warm DM with FSL $\lambda_{\rm FSL} < 0.1$ Mpc is still allowed, and can be favourable over cold DM of FSL $\lambda_{\rm FSL} < 0.01$ Mpc due to the small-scale structure problems associated with the latter \cite{Drewes:2016upu}. Dark matter free-streaming length can be estimated from matter power spectrum inferred from the Lyman-$\alpha$ forest data \cite{Croft:2000hs, Kim:2003qt, Viel:2005qj, Irsic:2017ixq} and also from Quasar data \cite{Hsueh:2019ynk}. Such estimates are also supported by theoretical and simulation based results \cite{Colombi:1995ze, Boyarsky:2008xj, deVega:2009ku, Schneider:2011yu}. For some recent discussions on structure formation constraints on DM production mechanisms, please see \cite{Merle:2013wta, Decant:2021mhj, Ballesteros:2020adh} and references therein. The detailed calculations related to FSL of DM are given in appendix \ref{appen_FSL}. 

\begin{figure}[h!]
\includegraphics[height=6cm,width=8.0cm,angle=0]{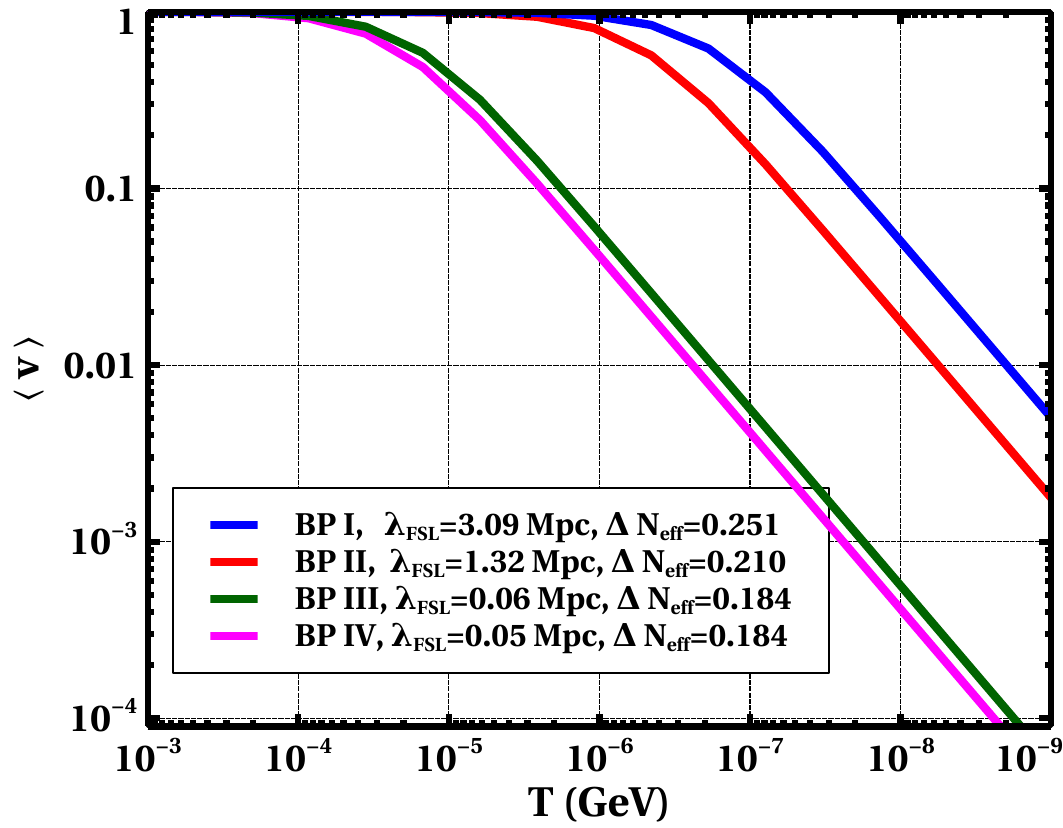}
\caption{Thermal average velocity of FIMP dark matter as a function of temperature for four different benchmark points.}
\label{fig:FSL}
\end{figure}

As shown in appendix \ref{appen_FSL}, the FSL of DM depends primarily on the production temperature, DM mass and the production mechanism of DM or distribution function of DM. A higher production temperature gives a smaller FSL due to high momentum redshift making DM non-relativistic earlier and vice-versa. A different production mechanism also gives different FSL. In our case, the DM is produced due to decay of a frozen  out scalar. Fig. \ref{fig:FSL} shows the average velocity of DM for four different benchmark parameters. The parameters are shown in table \ref{tab:case_2} along with total $\Delta N_{\rm eff}$ and FSL. The benchmark point (BP) II corresponds to a smaller $\phi_{1}$ mass than BP I. Hence, the freeze-out abundance of $\phi_{1}$ for BP II is smaller than that of BP I. This implies that a larger $\psi$ mass is required to satisfy DM abundance. As a result, we obtain a smaller FSL for BP II. However, for both BP I and BP II, the computed FSL keeps DM in hot DM category and hence ruled out from structure formation constraints. The FSL can be reduced by increasing the production temperature as well as increasing DM mass. BP III and BP IV have a larger $y_{\phi_{1}}$ giving a higher production temperature. Similarly it has a larger $\lambda_{H\phi_{1}}$ coupling, giving smaller $\phi_{1}$ freeze-out abundance. Hence for both BP III and BP IV, DM masses can be large while being consistent with relic abundance criteria. Combining these two effects, we get warm dark matter for BP III and BP IV. From the resulting on $\Delta N_{\rm eff}$, it can be seen that both BP III and BP IV remain within the sensitivity of future CMB experiments like CMB-S4.

\begin{table}[h!]
    \centering
    \begin{tabular}{|c|c|c|c|c|c|c|c|c|c|c|}
    \hline
    \multirow{2}{*}{} & \multicolumn{6}{|c|}{Parameters} & \multirow{2}{*}{$\Omega_{\rm DM} {\rm h}^2$} & \multirow{2}{*}{$\Delta {\rm N_{eff}}$} & \multirow{2}{*}{FSL(Mpc)}\\ \cline{2-7} \multirow{2}{*}{} &   $m_{\phi_1}$(GeV) & $\lambda_{H\phi_1}$ & $y_{\phi_1}$ & $m_{Z'}$(GeV) & $g_{BL}$ & $m_{\psi}$(keV)& \multirow{2}{*}{} & \multirow{2}{*}{} & \multirow{2}{*}{}\\ \hline 
    BP I & $1000$ & $5\times10^{-5}$ & $10^{-10}$ & $500$ & $0.001$ & $126$ & $0.12$ & $0.251$ & $3.09$ \\
    BP II & $500$ & $5\times10^{-5}$ & $10^{-10}$  & $500$ & $0.001$ & $233$ & $0.12$ & $0.210$ & $1.32$ \\
    BP III & $1000$ & $2\times10^{-4}$ & $10^{-9}$ & $500$ & $0.001$ & $970$ & $0.12$ & $0.184$ & $0.06$ \\
    BP IV & $500$ & $2\times10^{-4}$ & $10^{-9}$ & $500$ & $0.001$ & $938$ & $0.12$ & $0.184$ & $0.05$ \\
    \hline
    \end{tabular}
        \caption{Benchmark parameters for FIMP DM}
    \label{tab:case_2}
\end{table}

\begin{figure}[h!]
\includegraphics[scale=0.36]{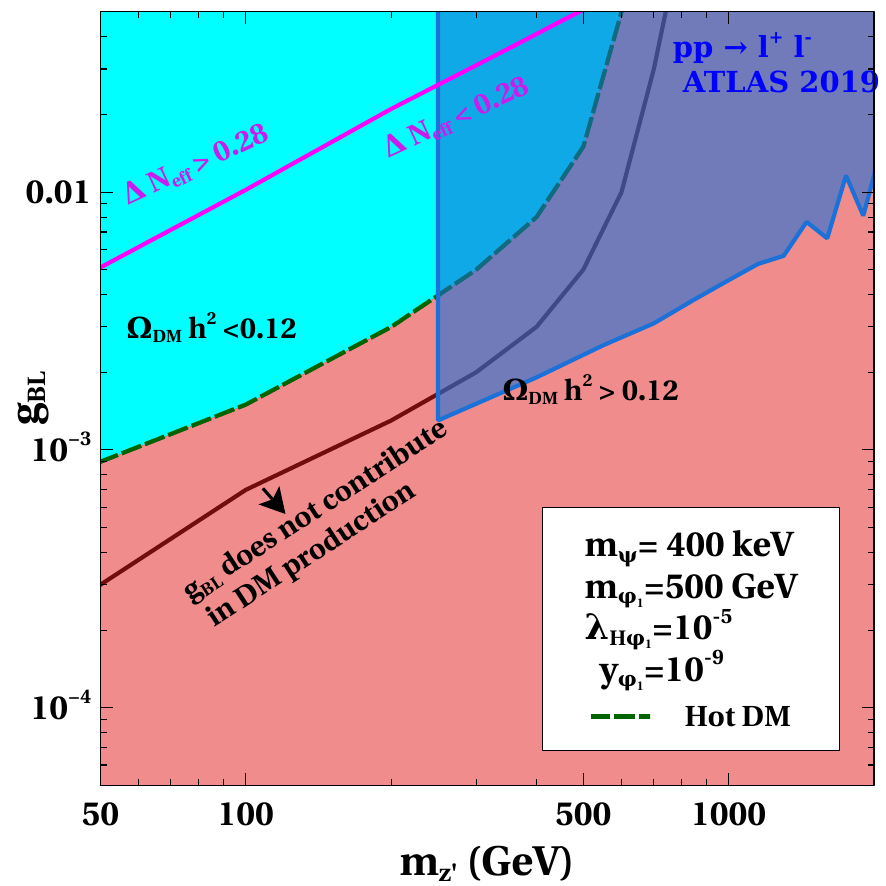}
\includegraphics[scale=0.36]{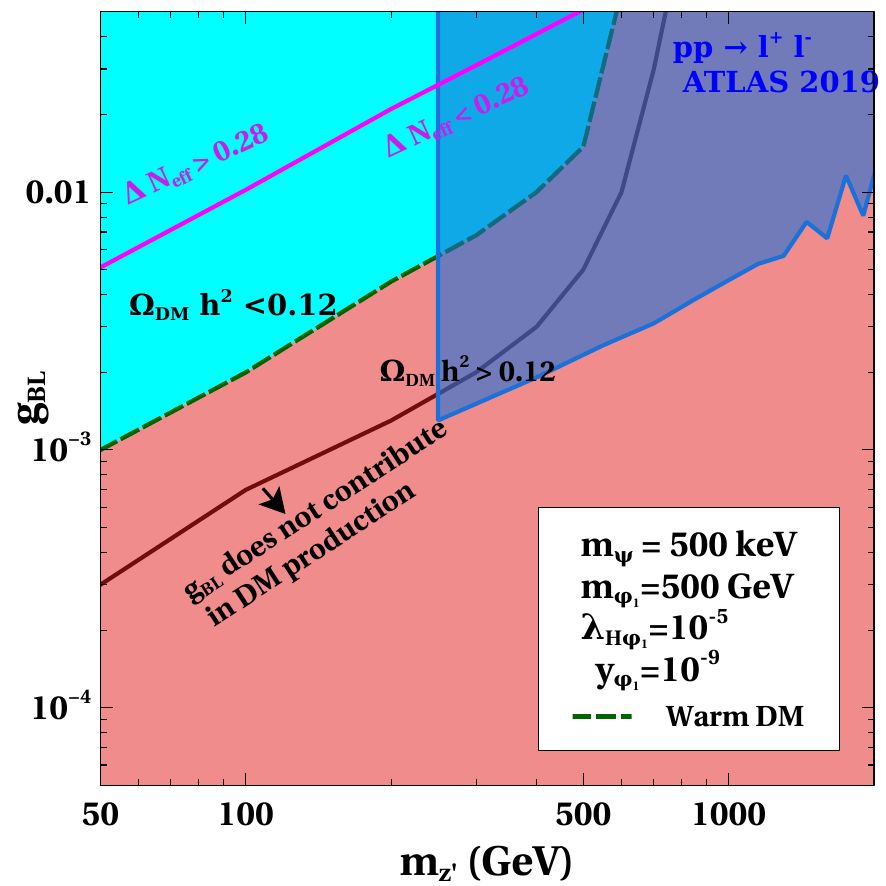}
\includegraphics[scale=0.36]{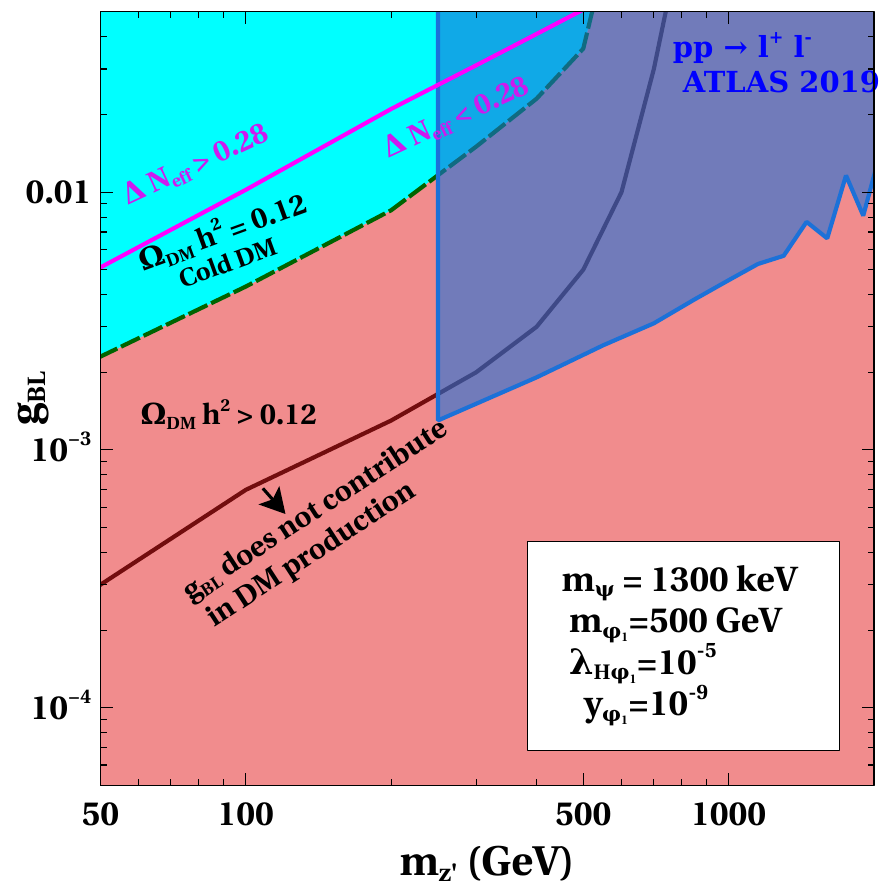}
\caption{Parameter space in $m_{Z'}$ versus $g_{BL}$ plane for the FIMP scenario considering three different values of DM mass. The other parameters are kept fixed as $m_{\phi_{1}}=500$ GeV, $\lambda_{H\phi_{1}}=10^{-5}$, $y_{\phi_{1}}=10^{-9}$.}
\label{fig:fimpbound}
\end{figure}

Fig. \ref{fig:fimpbound} shows the parameter space in $m_{Z'}$ versus $g_{BL}$ parameter space for the FIMP scenario. The left, middle and right panel plots correspond to dark matter mass of $400$ keV, $500$ keV and $1300$ keV respectively. The other relevant parameters are fixed as $m_{\phi_{1}}=500$ GeV, $\lambda_{H\phi_{1}}=10^{-5}$, $y_{\phi_{1}}=10^{-9}$. For the region below the solid brown line in each of these plots, $U(1)_{B-L}$-portal coupling does not play any role in FIMP DM production. In that region, the freeze-out abundance of $\phi_{1}$, responsible for DM production, is determined by the Higgs-portal coupling $\lambda_{H\phi_{1}}$. In other words, the cross section $\langle{\sigma v}\rangle _{_{\phi_{1} \phi_{1}^{\dagger{}} \to SM \overline{SM}}}$ dominates over $\langle{\sigma v}\rangle _{_{\phi_{1} \phi_{1}^{\dagger{}} \to Z' Z'}}$. In the region above the solid brown line, the freeze-out abundance of $\phi_{1}$ is determined by the $U(1)_{B-L}$-portal coupling. The dashed line in the left and middle panel plots, separating the pink and cyan shaded regions, represent the parameter space satisfying correct DM relic for $m_{\psi} = 400$ keV and $m_{\psi} = 500$ keV 
respectively. However, for DM mass $m_{\psi} = 400$ keV, the free-streaming length turns out to be very large $\sim 0.1$ Mpc, keeping it in hot DM ballpark and hence ruled out. In the middle panel plot, due to a higher DM mass $m_{\psi} = 500$ keV, the dashed line corresponds to an intermediate FSL keeping it in the warm DM regime which is still allowed by structure formation constraints. In the region below the dashed line, the freeze-out abundance of $\phi_{1}$ is more than that for the dashed line, leading to a larger FIMP DM abundance $\Omega_{\rm DM} h^2 > 0.12$, produced from $\phi_1$ decay. On the other hand, the region above the dashed line have $\Omega_{\rm DM} h^2 < 0.12$ due to smaller freeze-out abundance of $\phi_1$. This is understood from the fact that a larger $U(1)_{B-L}$-portal interaction leads to smaller freeze-out abundance of $\phi_1$ and vice versa. While the total $\Delta N_{\rm eff}$ can have both thermal and non-thermal contributions, for the choice of parameters in Fig. \ref{fig:fimpbound}, the non-thermal contribution is suppressed compared to the thermal contribution. The magenta solid line corresponds to $\Delta N_{\rm eff}=0.28$. The region above this line has $\Delta N_{\rm eff} > 0.28$ and hence ruled out from Planck 2018 limits. In the right panel plot, the dashed line corresponds to relic satisfying region for DM mass $m_\psi=1300$ keV. The region below the dashed line correspond to over-abundance like before. However, the region above the dashed line is no longer under-abundant, but satisfies the current DM abundance. For this region of parameter space and chosen DM mass, the production of FIMP DM while $\phi_1$ is in equilibrium dominates compared to the production after $\phi_1$ freeze-out. This can also be understood from the evolution plots shown in Fig. \ref{fig:FIMP_comparison}. 
From the left panel plot of Fig. \ref{fig:FIMP_comparison}, we see that there is a non-zero yield of DM $Y_{\psi}$ from the decay of $\phi_{1}$ when the latter is in equilibrium. Additionally, the freeze-out abundance of $\phi_{1}$ ($Y_{\phi_{1}} (x \rightarrow \infty)$) decreases as we increase $g_{BL}$. Combining these two, we can have a situation where for sufficiently large $U(1)_{B-L}$-portal couplings, $Y_{\psi}$ from the decay of $\phi_{1}$ in equilibrium is larger compared to the post freeze-out production. Hence, $U(1)_{B-L}$-portal couplings do not determine the comoving abundance of $Y_{\psi}$ and as a result, the region above the dashed line satisfies the correct DM relic in the right panel plot of Fig. \ref{fig:fimpbound}. In all the plots, the region below the magenta solid line has $0.28 > \Delta N_{\rm eff} > 0.14$ keeping it within reach of future CMB experiments like CMB-S4. The blue shaded region towards the upper right corner indicates the region ruled out from the large hadron collider (LHC) bounds, for specifically from ATLAS experiments at 13 TeV centre of mass energy \cite{ ATLAS:2019erb, ATLAS:2017fih}. Similar bounds also exist from the CMS experiment at the LHC \cite{Sirunyan:2018exx}. {The details of the LHC bound on $g_{BL}-m_{Z'}$ plane is given in appendix \ref{appen_gBL_mz'}.} A relatively weaker bound exists from the large electron positron (LEP) collider disfavouring the region $m_{Z'}/g_{BL} < 7$ TeV \cite{Carena:2004xs, Cacciapaglia:2006pk}. The parameter space shown in Fig. \ref{fig:fimpbound} already satisfies the LEP bound.


\subsection*{When $\phi_{1}$ is always in equilibrium:}
Before moving onto the WIMP DM scenario, we briefly comment on the possibility of FIMP DM production from $\phi_1$ when the latter remains in equilibrium throughout the production.
When $\phi_{1}$ is in equilibrium during the production of DM, the comoving number density of $\psi$ and comoving energy density of $\nu_{R}$, respectively, are given as
\begin{equation} \label{BE_eq}
     \frac{dY_{\psi}}{dx} = \frac{ \beta }{x \mathcal{H} } \Gamma_{\phi_{1}} \frac{K_1(x)}{K_2(x)} Y^{\rm eq}_{\phi_{1}},
\end{equation}
\begin{equation}
       \frac{d \widetilde{Y}}{dx} = \frac{\beta}{\mathcal{H} s^{1/3} x} \langle E\Gamma \rangle _{\phi_{1}}  Y^{\rm eq}_{\phi_{1}}.
       \label{1nur_freezed_out}
\end{equation}
These two equations can be analytically solved to get the asymptotic abundances (in the limit $m_{\psi} \ll m_{\phi_{1}}$)
\begin{eqnarray}\label{BE1_eq}
    Y_{\psi} (x \rightarrow \infty) &=& \frac{135\,}{16\pi \times 1.66 \times 
    8 \pi^3 g^{s}_{*} \sqrt{g^{\rho}_{*}}} M_{\rm pl} \,
   \frac{y^2_{\phi_{1}}}{m_{\phi_1}}\,,
   \nonumber \\   
    \widetilde{Y} (x \rightarrow \infty) &=& \frac{675\,}{32\pi \times 1.66\times 8\pi^3 g^{s}_{*}\sqrt{g^{\rho}_{*}}} \left(\frac{45}{2\pi^2\,g^{s}_{*}}\right)^{1/3} M_{\rm pl}\,
    \frac{y^2_{\phi_{1}}}{m_{\phi_1}}\,,
\end{eqnarray}
where $M_{\rm pl}$ is the Planck mass. With this, the DM abundance and contribution to extra radiation energy density can be calculated as -
\begin{eqnarray}\label{abun_eq}
    \Omega_{\rm DM} h^2 &=& 2 \frac{m_{\psi} s^{0}}{\rho^{0}_{c}} Y_{\psi} (x \rightarrow \infty) h^2 \nonumber \\
    \Delta N^{\rm non-th}_{\rm eff} &=& 2\times 3 \times \left(\frac{s^{4/3}}{\rho_{\nu_{L}}}\right)_{\rm T= 10\, MeV} \widetilde{Y} (x \rightarrow \infty) \simeq 0.61 \widetilde{Y} (x \rightarrow \infty),
\end{eqnarray}
where $s^{0}$ and $\rho^{0}_{c}$ denote the entropy density and critical energy density, respectively, at the present epoch. $\rho_{\nu_{L}}$ denotes energy density of one species of SM neutrino and $h$ represents $H_{0}/100$ with $H_{0}$ being the current expansion rate of the universe. Here, $m_{\phi_{1}}, y_{\phi_{1}}$ and $m_{\psi}$ determine the DM abundance and $\Delta N^{\rm non-th}_{\rm eff}$ is determined by $m_{\phi_{1}}$ and $y_{\phi_{1}}$ only. The thermal contribution to $\Delta N_{\rm eff}$ is determined by the parameters $m_{Z'}$ and $g_{BL}$. When DM abundance is satisfied, it turns out that the non-thermal contribution to $\Delta N_{\rm eff}$ is way below the minimum thermal contribution. So, the total $\Delta N_{\rm eff}$ is only determined by the thermal contribution (shown in Fig. \ref{fig:thermal_Neff}). 

\begin{figure}[h!]
\includegraphics[height=5.5cm,width=7.5cm,angle=0]{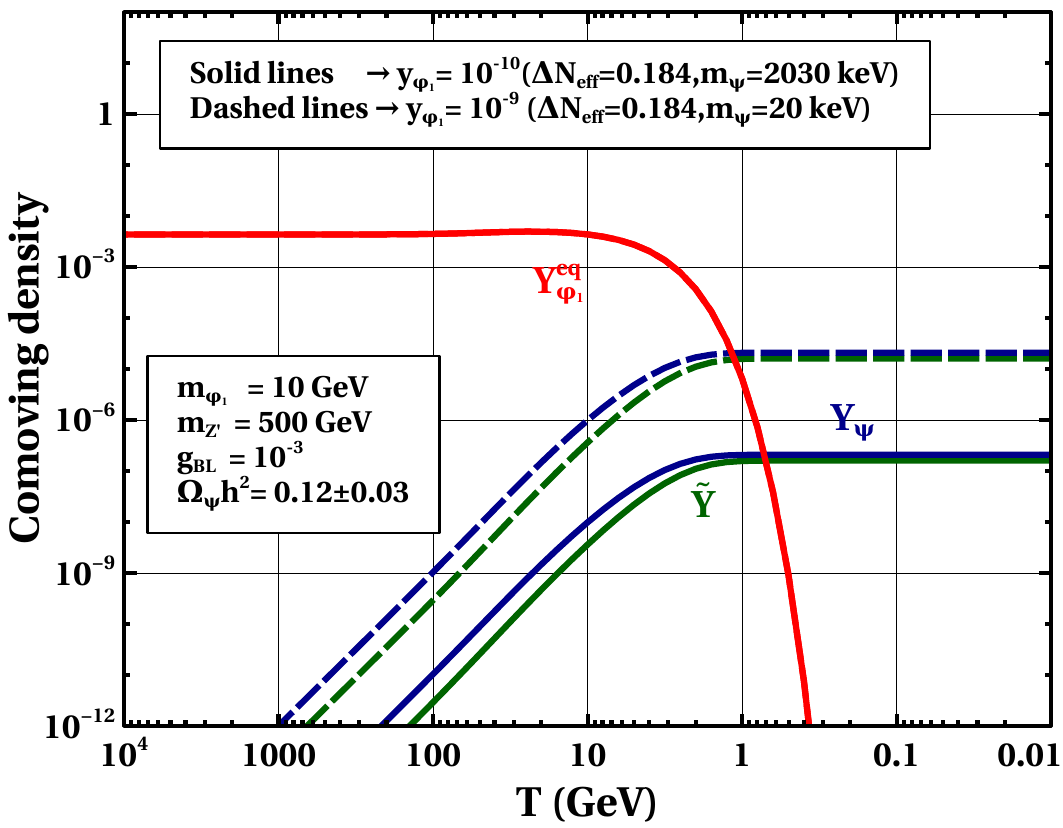}
\includegraphics[height=5.5cm,width=7.5cm,angle=0]{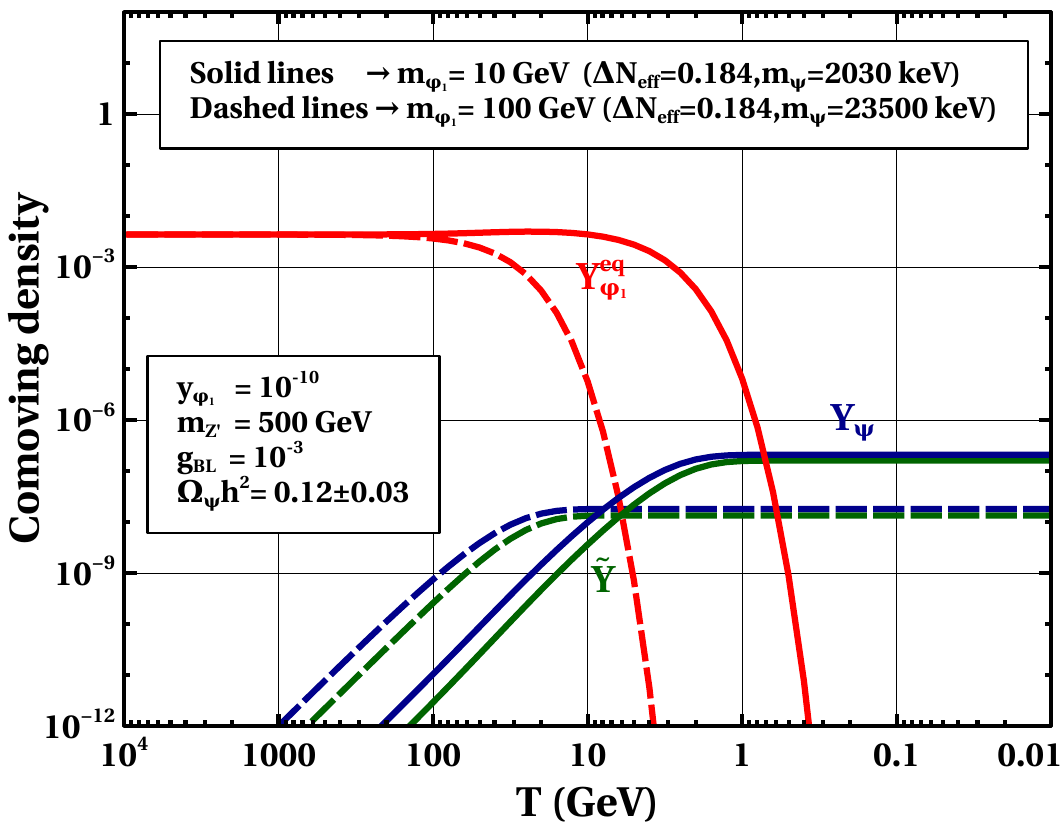}
\caption{Evolution of comoving number densities of $\phi_{1}$ and $\psi$ and comoving energy density of $\nu_{R}$ for two different values of $y_{\phi_{1}}$ (left panel) and $m_{\phi_{1}}$ (right panel).}
\label{fig:FIMP_eq_comparison}
\end{figure}

\begin{table}[h!]
    \centering
    \begin{tabular}{|c|c|c|c|c|c|c|c|c|c|}
    \hline
    \multirow{2}{*}{} & \multicolumn{5}{|c|}{Parameters} & \multirow{2}{*}{$\Omega_{\rm DM} {\rm h}^2$} & \multirow{2}{*}{$\Delta {\rm N_{eff}}$} & \multirow{2}{*}{FSL(Mpc)}\\ \cline{2-6} \multirow{2}{*}{} &   $m_{\phi_1}$(GeV) & $y_{\phi_1}$ & $m_{Z'}$(GeV) & $g_{BL}$ & $m_{\psi}$(keV)& \multirow{2}{*}{} & \multirow{2}{*}{} & \multirow{2}{*}{}\\ \hline 
    BP I & $10$  & $10^{-10}$ & $500$ & $0.001$ & $2030$ & $0.12$ & $0.184$ & $0.0009$ \\
    BP II & $10$ &  $10^{-9}$  & $500$ & $0.001$ & $20$ & $0.12$ & $0.184$ & $0.06$ \\
    BP III & $100$  & $10^{-10}$ & $500$ & $0.001$ & $23500$ & $0.12$ & $0.184$ & $0.0001$ \\
    BP IV & $100$  & $10^{-9}$ & $500$ & $0.001$ & $235$ & $0.12$ & $0.184$ & $0.0064$ \\
    \hline
    \end{tabular}
        \caption{Table for FIMP DM (equilibrium $\phi_{1}$)}
    \label{tab:case_eq}
\end{table}

Fig. \ref{fig:FIMP_eq_comparison} shows evolutions of comoving abundances of $\phi_{1}$, $\psi$ and $\nu_{R}$ for two different values of $y_{\phi_{1}}$ (left panel) and $m_{\phi_{1}}$ (right panel). For $m_{\psi} \ll m_{\phi_{1}}$ the evolutions for comoving number density of $\psi$ and comoving energy density of $\nu_{R}$ are independent of $m_{\psi}$. So in all the evolution plots, the dark matter masses are chosen in such a way that observational DM abundance is satisfied. The solid lines in the left plot have $y_{\phi_1} = 10^{-10}$ and DM mass $m_{\psi} = 2030$ keV. The non-thermal contribution to $\Delta N_{\rm eff}$ from frozen-in $\nu_{R}$ is $6.5 \times 10^{-6}$. For the dashed lines, the Yukawa coupling is larger by a factor of 10. Consequently, we have higher $Y_{\psi} (x\rightarrow \infty)$ and $\tilde{Y} (x\rightarrow \infty)$. From Eq. \eqref{abun_eq}, we get DM mass to be $20$ keV and $\Delta N^{\rm non-th}_{\rm eff} = 6.5 \times 10^{-4}$. The solid lines in the right plot have the same parameters as those of in the left panel plot. The dashed lines in the right panel plot have $m_{\phi_{1}} = 100$ GeV. Due to higher mass of $\phi_{1}$, the Boltzmann suppression occurs early on, leading to smaller $Y_{\psi} (x\rightarrow \infty)$ and $\tilde{Y} (x\rightarrow \infty)$. As a result DM mass for the solid lines is about $23.5$ MeV and $\Delta N^{\rm non-th}_{\rm eff} = 5.5 \times 10^{-7}$. We have kept $B-L$ gauge coupling and mass fixed for all the plots, $g_{BL}=10^{-3}$ and $m_{Z'}=500$ GeV. This gives a thermal contribution to $\Delta N_{\rm eff} = 0.184$. As in all the plots non-thermal contribution is way smaller than the thermal counterpart, so the total $\Delta N_{\rm eff}$ is determined by the thermal contribution as shown in the plot. Since $U(1)_{B-L}$-portal parameters do not decide DM abundance in this case, we do not show any summary plot like before.

Table \ref{tab:case_eq} shows four benchmark points (BP) and their corresponding contribution to effective number of relativistic species and structure formation. BP I and BP II have same $m_{\phi_{1}} = 10 $ GeV  whereas BP III and BP IV have $m_{\phi_{1}} = 100$ GeV. As the Yukawa coupling is ten times larger in BP II compared to BP I, the corresponding DM mass is 100 times smaller in BP II (see Eq. \eqref{BE1_eq}). The similar trend is followed between BP III and BP IV. Due to smaller DM mass, $\lambda_{\rm FSL}$ is larger in both the BP II (compared to BP I) and BP IV (compared to BP III). Again from equation \eqref{abun_eq}, we get that $m_{\phi_{1}}$ and $m_{\psi}$ are correlated for constant DM abundance. Hence a larger $m_{\phi_{1}}$ gives a larger $m_{\psi}$ (BP III, BP IV vs BP I, BP II). The DM production temperature ($T \sim m_{\phi_{1}}$) is also larger for larger $m_{\psi_{1}}$. Combining these two, we get a smaller $\lambda_{\rm FSL}$ for BP III (BP IV) compared to BP I (BP II). Among the four benchmark points mentioned in the table, only BP II gives warm DM whereas rest points give cold DM. As all the points have same $B-L$ gauge coupling and mass, we have same $\Delta N_{\rm eff}$.

\section{WIMP type dark matter}
\label{sec:wimp}
\begin{figure}
    \includegraphics[scale=0.5]{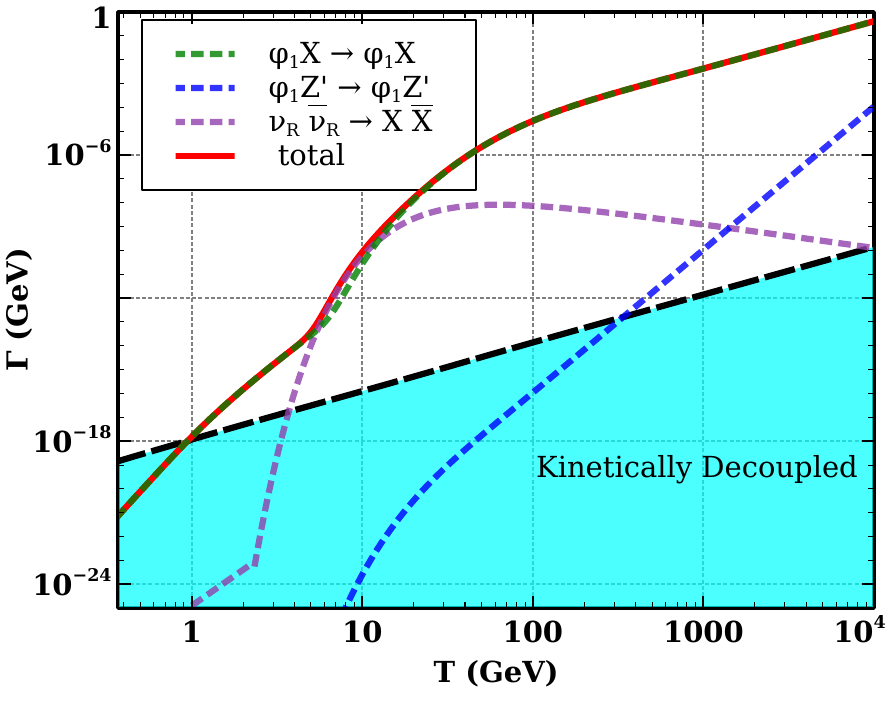}
     \includegraphics[scale=0.5]{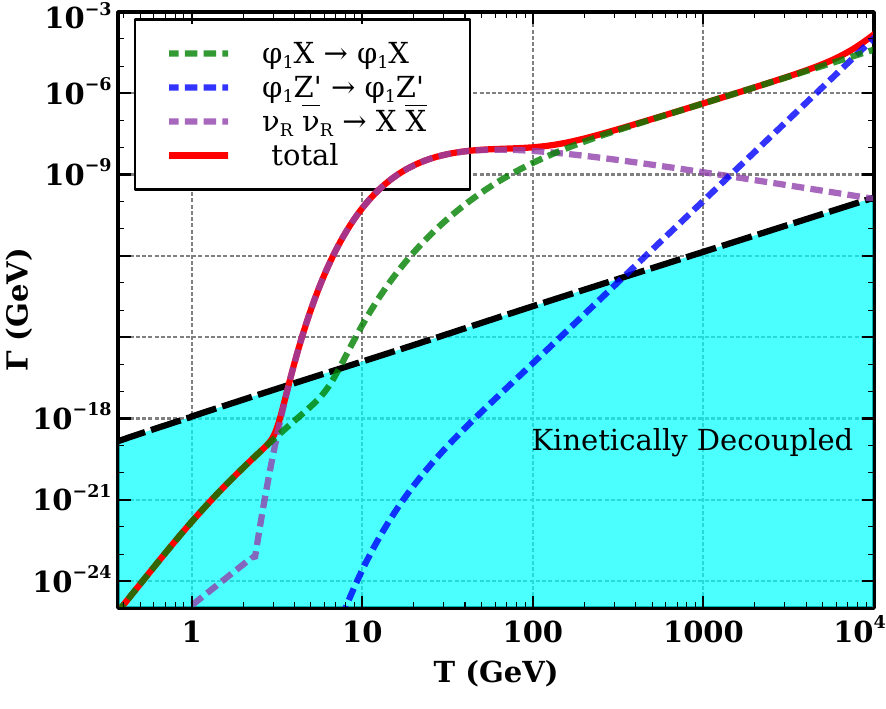}
    \caption{Interaction rate as a function of temperature for processes responsible for keeping dark sector in equilibrium with the SM bath. The benchmark parameters chosen are $\lambda_{H\phi_{1}} =10^{-2} \, (\rm left \, panel), 10^{-4} \, (\rm right \, panel)$, $m_{\phi_{1}} = 100$ GeV, $m_{Z'} = 100$ GeV, $g_{BL}=10^{-4}$.}
    \label{fig:interaction_rate}
\end{figure}

If the Yukawa coupling among the dark sector particles i.e. $\phi_{1}$, $\psi$ and $\nu_{R}$ is sufficiently strong ($y_{\phi_{1}} \gg 10^{-7}$), the dark sector can be in thermal equilibrium among themselves even after their decoupling from the SM bath. Since the thermalisation of $\phi_1, \nu_R$ with the SM bath relies on Higgs portal and $U(1)_{B-L}$ portal couplings instead of Yukawa $y_{\phi_{1}}$, we have two different sub-cases depending upon whether the Yukawa interactions in dark sector go out of equilibrium before or after the dark sector decouples from the thermal bath. Among the dark sector particles, $\phi_{1}$ and $\nu_{R}$ can have interactions with other particles in the bath via the processes - $\phi_{1} X \to \phi_{1} X$, $\phi_{1} Z' \to \phi_{1} Z'$ and $\nu_{R} X \to \nu_R X$, here $X$ denotes SM particles. Since $Z'$ has sizeable interactions with all the SM fermions, we consider it to be in the bath while dark sector particles decouple. While the first two processes can arise via contact interactions too, the $\nu_R$ scattering arises via mediation of heavy $Z'$ boson only. Unlike other interactions, $\nu_R$ interactions with the bath rely significantly on resonant enhancement and hence we consider $\nu_R \bar{\nu}_R \rightarrow X \bar{X}$ via resonantly enhanced s-channel mediation of $Z'$ to calculate its decoupling \cite{Abazajian:2019oqj}. Therefore, the ratio of interaction rate of dark sectors with SM bath to the expansion rate, $\mathcal{H}$, can be written as \cite{Gondolo:2012vh, Heeck:2014zfa}
\begin{eqnarray}
     \frac{\Gamma_{\rm total}}{\mathcal{H}} = \frac{1}{\mathcal{H}} \left[n^{\rm eq}_{X} \left(\langle \sigma v \rangle_{\phi_{1}X \to \phi_{1} X} + \langle \sigma v \rangle_{\phi_{1}Z' \to \phi_{1} Z'}\right) + n^{\rm eq}_{\nu_{R}} \langle \sigma v \rangle_{\nu_{R} \bar{\nu}_{R} \to X \bar{X}} \right].
\end{eqnarray}
 The decoupling temperature of dark sector is approximately calculated by comparing the total interaction rate with the Hubble expansion rate i.e. $\frac{\Gamma_{\rm total}}{\mathcal{H}} \approx 1$. Fig. \ref{fig:interaction_rate} shows interaction rates of various processes responsible for keeping the dark sector in bath with the SM as a function of temperature. In the left panel plot, a large scalar portal coupling $\lambda_{H\phi} = 10^{-2}$ dictates the decoupling temperature whereas in the right panel plot, the resonantly enhanced process $\nu_{R} \bar{\nu}_{R} \to X \bar{X}$ determines the decoupling temperature due to smaller scalar portal coupling $\lambda_{H\phi} = 10^{-4}$.

\begin{figure}
    \includegraphics[scale=0.35]{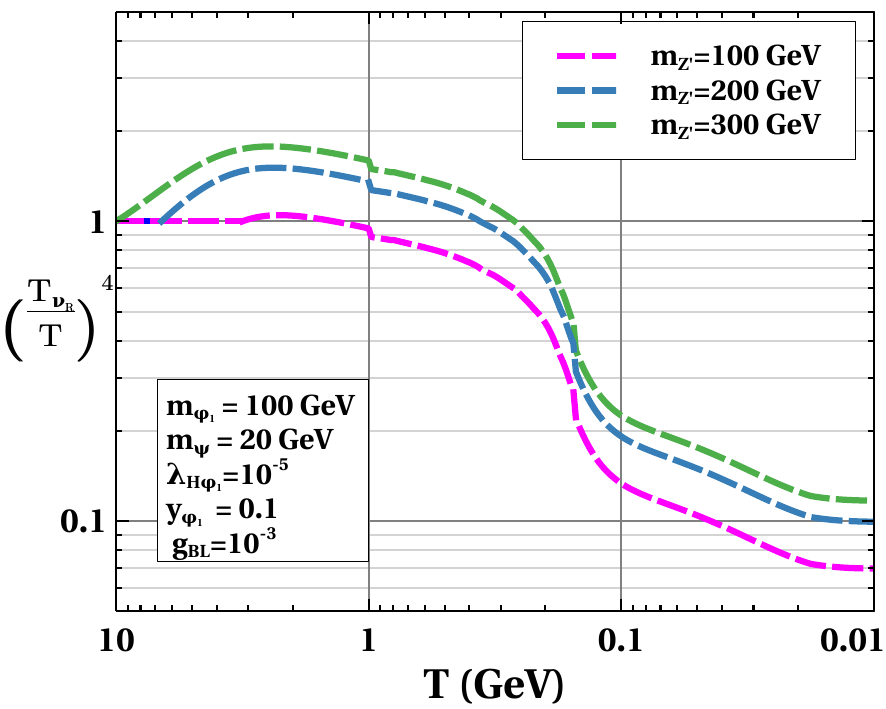}
    \includegraphics[scale=0.35]{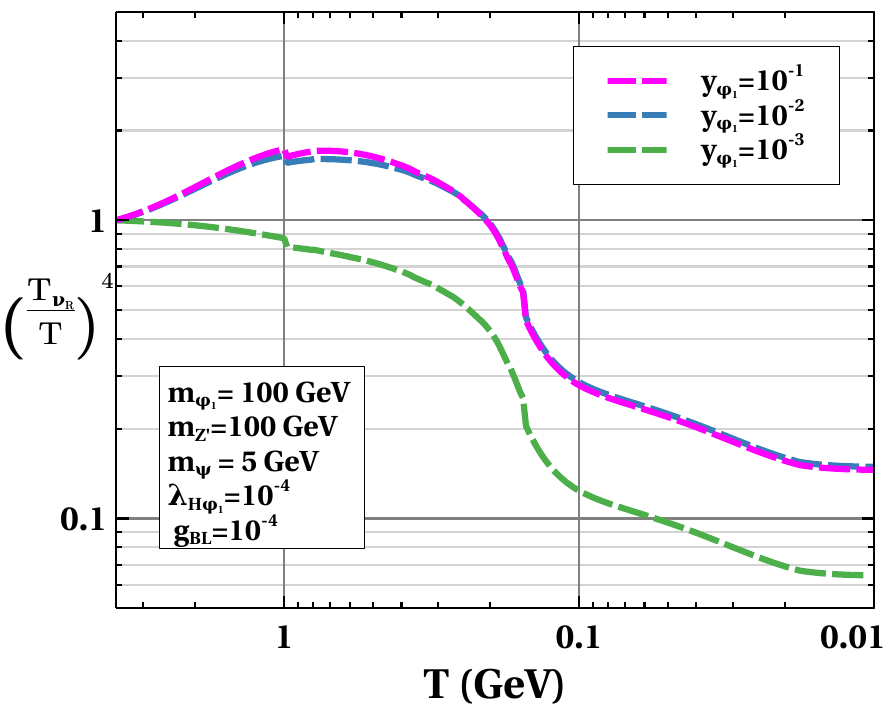}
    \includegraphics[scale=0.35]{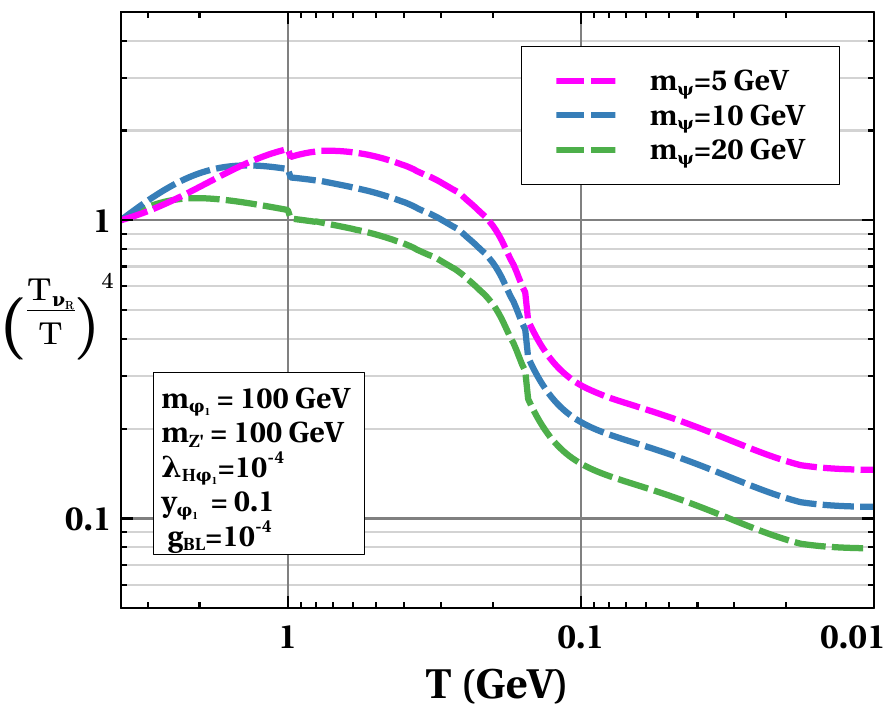}

    \caption{Evolution of dark sector temperature after its decoupling from SM bath ($T<T_{\rm dec}$) for different values of $m_{Z'}$ (left panel), dark sector Yukawa coupling $y_{\phi_{1}}$ (middle panel) and dark matter mass $m_{\psi}$ (right panel). The rest of the parameters are kept fixed and are shown in respective plots.}
    \label{fig:LY_neff}
\end{figure}

 To track the evolution of dark sector, one has solve the Boltzmann equation in terms of comoving number densities $Y_{\phi_{1}}$ and $Y_{\psi}$. Using the approximation $\frac{Y_{i}}{Y}\approx \frac{Y^{\rm eq}_{i}}{Y^{\rm eq}}$ ($i=\phi_{1},\psi$), with $Y=Y_{\phi_{1}}+Y_{\psi}$, we can write a single Boltzmann equation as 
 \begin{eqnarray}\label{before_Tdec}
  \frac{dY}{dx} = \frac{1}{2}\frac{\beta s}{\mathcal{H} x} \langle \sigma v\rangle_{\rm eff}  \left((Y^{\rm eq})^2-Y^2\right),
\end{eqnarray}
where
\begin{eqnarray}
    \langle \sigma v\rangle_{\rm eff} = \frac{(Y^{eq}_{\phi_{1}})^2\langle \sigma v\rangle_{\phi_{1}\phi^{\dagger}_{1} \to X\bar{X},\nu_{R}\Bar{\nu}_{R},Z'Z'} + (Y^{eq}_{\psi})^2\langle \sigma v\rangle_{ \psi\Bar{\psi} \to \nu_{R}\Bar{\nu}_{R}}}{(Y^{eq}_{\phi_{1}}+Y^{eq}_{\psi})^2}.
\end{eqnarray}
The equation \eqref{before_Tdec} can be solved numerically from a sufficiently high temperature to the decoupling temperature, $T_{\rm dec}$. If the Yukawa coupling $y_{\phi_{1}}$ is large enough, even after $T< T_{\rm dec}$, the dark sector particles $\phi_{1}$, $\psi$ and $\nu_{R}$ can be in dark sector equilibrium among themselves. We can track the dark sector temperature $T_{\nu_{R}}$($ \equiv T_{\rm DS}$) by solving the following equation.
\begin{eqnarray}\label{after_Tdec}
  \frac{d\xi}{dx} &=& \frac{1}{x} \left(-\frac{1}{2}\frac{\beta x^4 s^2}{4 \alpha \xi^3 H m^4_{\phi_{1}}} \langle E\sigma v\rangle_{\rm eff}\left((Y^{\rm eq})^2-Y^2\right) - (\beta -1)\xi \right),
\end{eqnarray}
where
\begin{eqnarray}
    \langle E\sigma v\rangle_{\rm eff} = \frac{(Y^{\rm eq}_{\nu_{R}})^2 \langle E\sigma v\rangle_{\nu_{R}\Bar{\nu}_{R}\to \psi \Bar{\psi}} + (Y^{\rm eq}_{\nu_{R}})^2 \langle E\sigma v\rangle_{\nu_{R}\Bar{\nu}_{R}\to \phi_{1} \phi^{\dagger}_{1}}}{(Y^{\rm eq}_{\phi_{1}}+Y^{\rm eq}_{\psi})^2},
\end{eqnarray}
and $\xi$ is the ratio of dark sector temperature to that of the SM bath, $\frac{T_{\nu_{R}}}{T}$ and $\alpha=6\times \frac{7}{8}\frac{\pi^2}{30}$. After decoupling, $\langle \sigma v\rangle_{\rm eff}$ in Eq. \eqref{before_Tdec} can depend upon both $x$ as well as $\xi$. We solve Eq. \eqref{before_Tdec} and Eq. \eqref{after_Tdec} simultaneously from $T=T_{\rm dec}$ to the onset of the BBN era namely, $T\sim 10$ MeV and estimate the corresponding DM relic and $\Delta N_{\rm eff}$. Fig. $\ref{fig:LY_neff}$ shows the evolution of $T_{\nu_{R}}$ for different benchmark parameters. Different $T_{\nu_{R}}$ can lead to changes in $\Delta N_{\rm eff}$. The first term on the RHS of Eq. \eqref{after_Tdec} is a positive term as $Y \geq Y^{\rm eq}$ for $T < T_{\rm dec}$. Thus, this term indicates the increase in $T_{\nu_{R}}$ over $T$ due to annihilation of heavy dark sector particles $\phi_{1}$ and $\psi$ into $\nu_{R}$. On the other hand, the second term on the RHS of the same equation indicates the decrease in $T_{\nu_{R}}$ over $T$ due to changes in entropy degrees of freedom $g^{s}_{*}$. The combined effects of these two terms can be seen in Fig. \ref{fig:LY_neff}. For an analytical description of the behaviour of the plots shown in Fig. \ref{fig:LY_neff}, please refer to appendix \ref{appen_dark_temp}. 


The left panel plot in Fig. \ref{fig:LY_neff} shows the variation of $\xi^4$ with respect to bath temperature $T$ for different choices of $m_{Z'}$. The rest of the parameter are kept fixed as shown in the plot. Increasing $m_{Z'}$ gives a higher decoupling temperature, $T_{\rm dec}$. As long as $T_{\rm dec} \lesssim m_{\phi_{1}}$, a higher decoupling temperature leads to a larger conversion of $\phi_{1}$ and $\psi$ to $\nu_{R}$ resulting in a larger $\frac{T_{\nu_{R}}}{T}$. Another way to understand it is from entropy conservation (see appendix \ref{appen_dark_temp}). For a higher $T_{\rm dec}$, we have a larger dark sector entropy degrees of freedom $g^{\rm s,DS}_{*}$ resulting in a larger $\frac{T_{\nu_{R}}}{T}$. From an analytical estimation, we get a maximum value of $\xi^4$ to be 2.6 provided $g^{s}_{*}$ remains same from the epoch of dark sector decoupling to DM freeze-out, as shown in appendix \ref{appen_dark_temp}. This is consistent with the numerical results shown in left panel of Fig. \ref{fig:LY_neff}. The reason why maximum value of $\xi^4$ has not reached $2.6$ in the figure is that $g^{s}_{*}$ changes from the period of kinetic decoupling of dark sector to DM freeze-out. The middle panel plot shows the results for different values of dark sector Yukawa coupling. As the dark sector Yukawa coupling does not affect the DM-SM decoupling temperature, so all the coloured lines correspond to the same decoupling temperature. For a larger Yukawa coupling, conversion of $\phi_{1}$ and $\psi$ to $\nu_{R}$ is possible and as a result, temperature of $\nu_{R}$ with respect to SM bath increases. For the green line in the middle panel plot of Fig. \ref{fig:LY_neff}, the dark sector freeze-out occurs before $T_{\rm dec}$ due to small Yukawa coupling. Hence the first term on the RHS of Eq. \eqref{after_Tdec} ceases and due to the second term, the ratio $\frac{T_{\nu_{R}}}{T}$ decreases. To see the effect of $m_{\psi}$, in the right panel plot, we vary $m_{\psi}$ keeping the other parameter constant. For a smaller $m_{\psi}$, freeze-out occurs late, hence we get a higher $\frac{T_{\nu_{R}}}{T}$ ratio.

\begin{figure}[h!]

\includegraphics[scale=0.35]{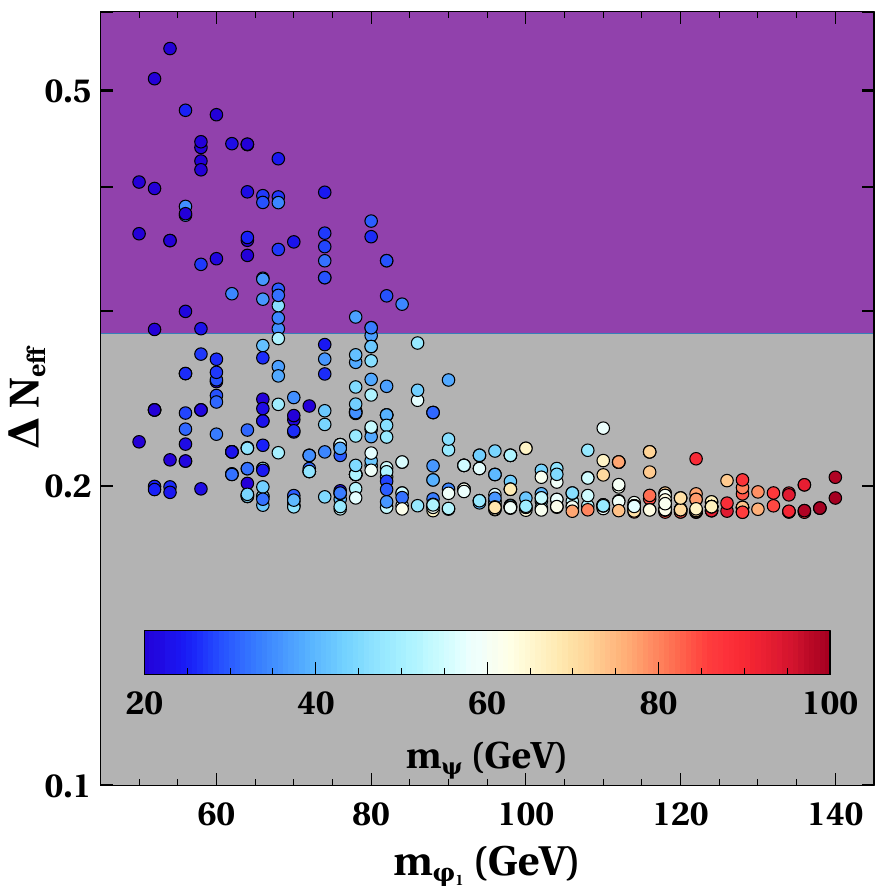}
\includegraphics[scale=0.35]{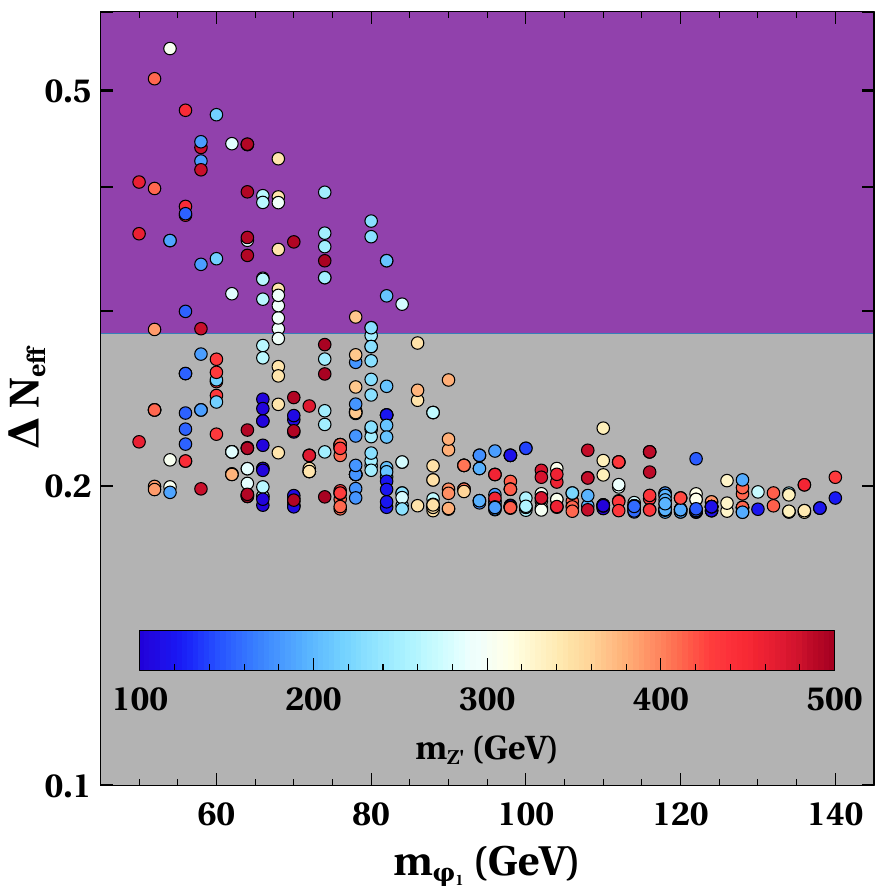}
\includegraphics[scale=0.35]{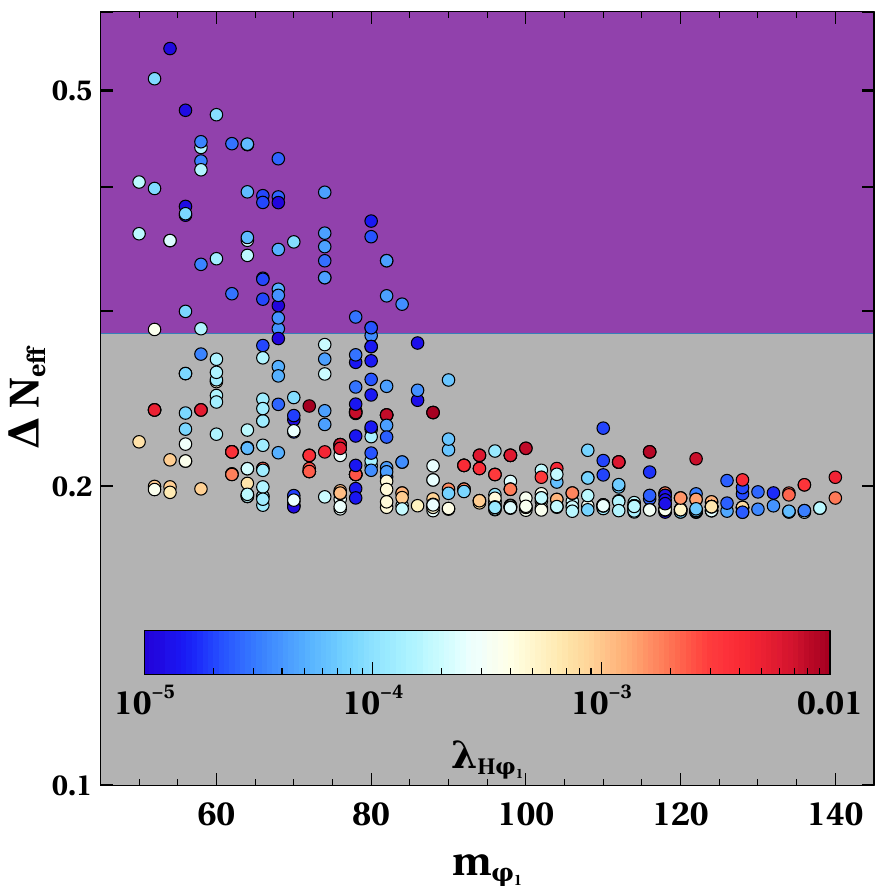}
\caption{Scan plot showing $\Delta N_{\rm eff}$ vs $m_{\phi_{1}}$ for different dark matter mass $m_{\psi}$ and different $m_{Z'}$ and $g_{BL}$.}
\label{fig:LY_scan}
\end{figure}

To see the complete picture in WIMP dark matter scenario, we perform a numerical scan by varying the following parameters as

\begin{eqnarray}\label{WIMP_parameters}
    50 \text{GeV} &< m_{\phi_{1}} < 150 \text{GeV}, \hspace{1 cm} 100 \text{GeV} &< m_{Z'} < 500 \text{GeV}, \\ \nonumber
    10 \text{GeV} &< m_{\psi} < 100 \text{GeV}, \hspace{1.5 cm} 10^{-5} &< \lambda_{H\phi_{1}} <10^{-2}, \\ \nonumber
    10^{-5} &< g_{BL} < 10^{-3}, \hspace{2 cm} 0.2 &< y_{\phi_{1}} < 0.3.
\end{eqnarray}

The corresponding results are shown in Fig. \ref{fig:LY_scan} with y-axis representing $\Delta N_{\rm eff}$ whereas x-axis showing the mass of $\phi_{1}$. The relevant parameters $m_{\psi}$, $m_{Z'}$ and $g_{BL}$ are shown in colour bars of left, middle and right panel plots respectively of Fig. \ref{fig:LY_scan}. All the points shown in this figure satisfy the requirements of correct WIMP DM relic abundance. The magenta shaded region denotes the region excluded by Planck 2018 bounds at $2\sigma$ CL while the grey shaded region remains within the reach of future experiments like CMB-S4. From the left panel plot, we can see that for some points with dark matter mass $m_{\psi} \lesssim 50$ GeV are already excluded by Planck 2018 data at $2\sigma$ CL. Similarly, from the middle and right panel plots, we get some points having $m_{Z'}$ between $100$ GeV and $500$ GeV and $g_{BL}$ between $10^{-3}$ and $10^{-5}$ GeV, that are already excluded. All the others points in these plots can be probed in the future CMB experiments like CMB-S4 keeping the detection prospects promising.


\begin{figure}[h!]
\includegraphics[height=7cm,width=7.0cm,angle=0]{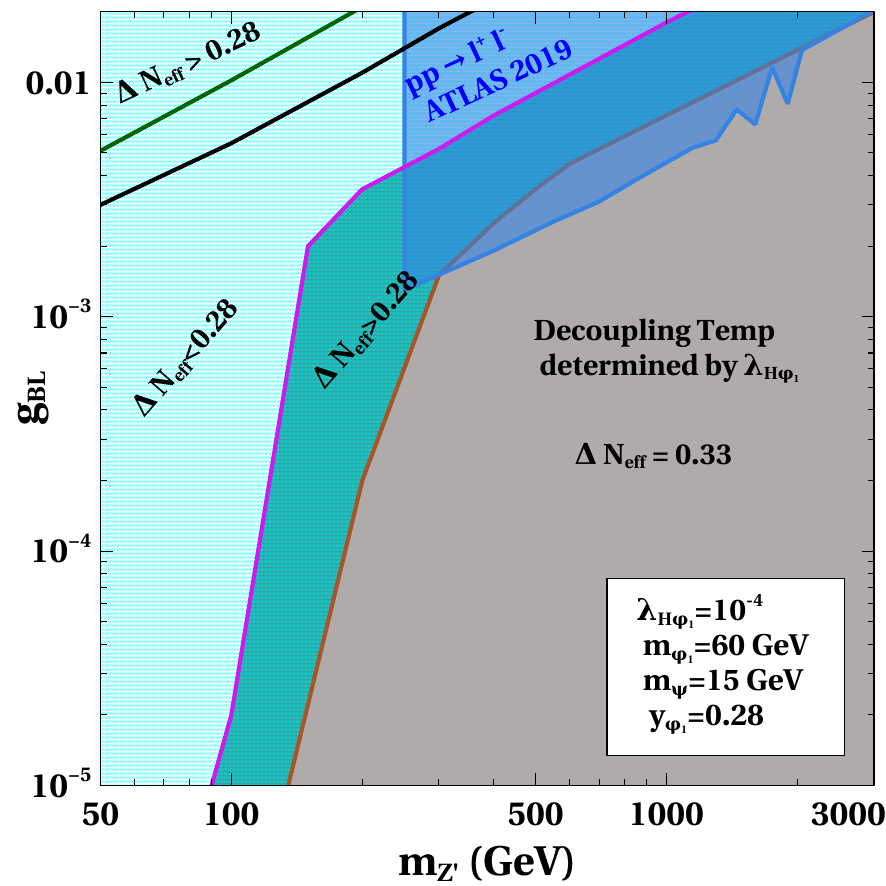}
\includegraphics[height=7cm,width=7.0cm,angle=0]{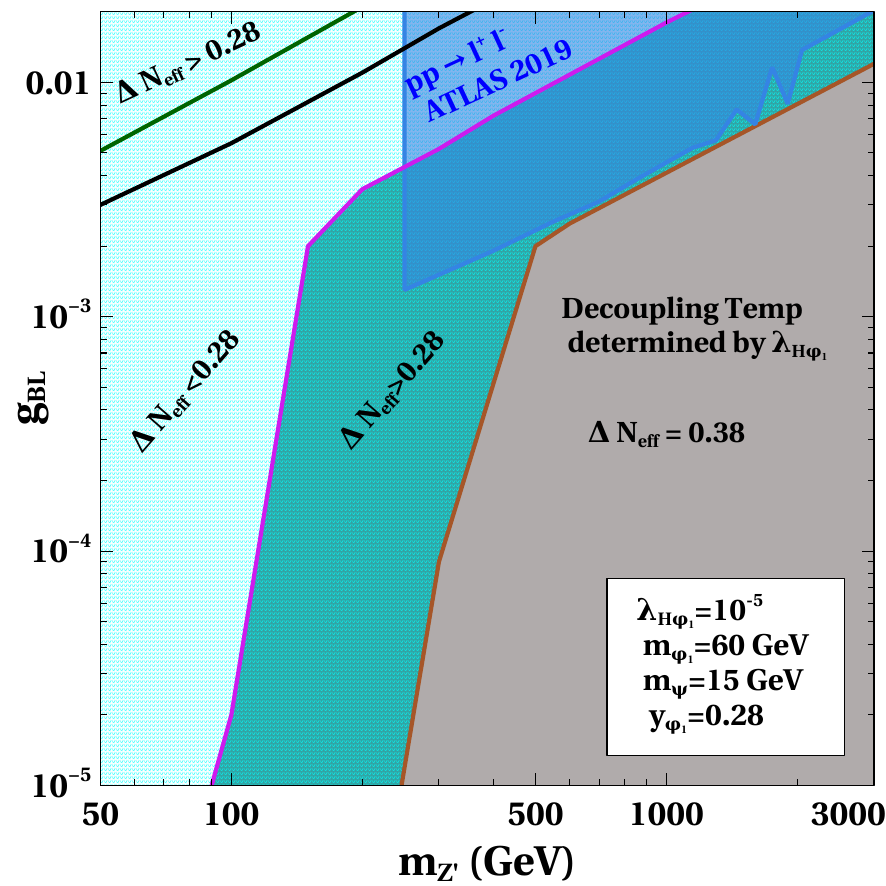}
\includegraphics[height=7cm,width=7.0cm,angle=0]{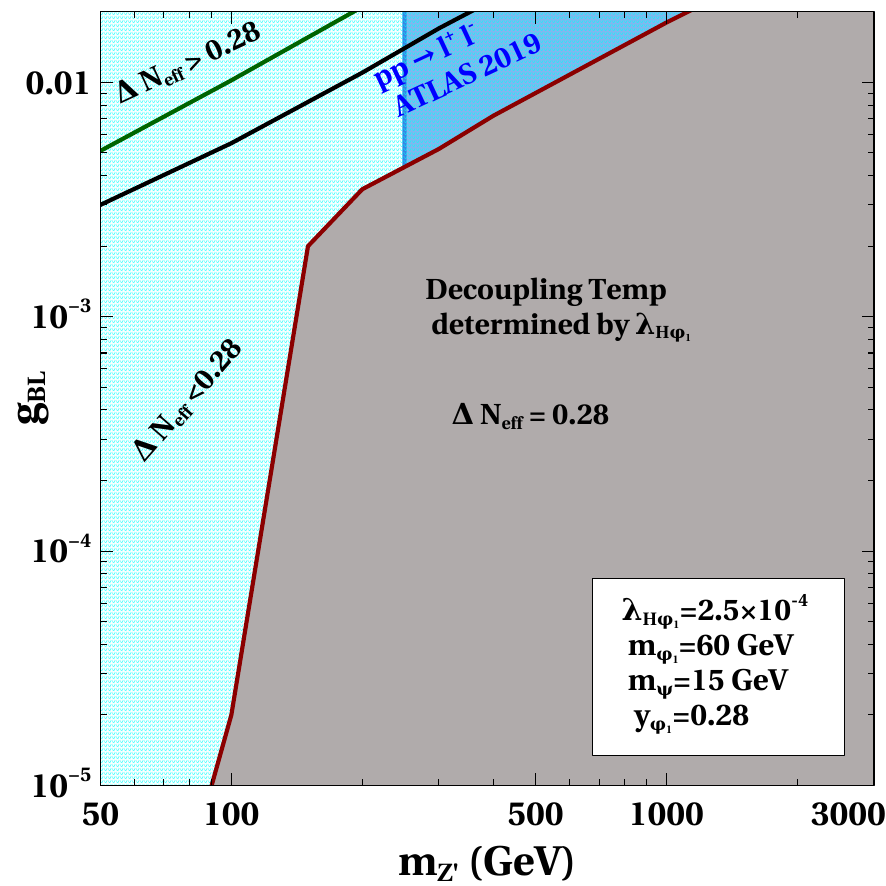}
\includegraphics[height=7cm,width=7.0cm,angle=0]{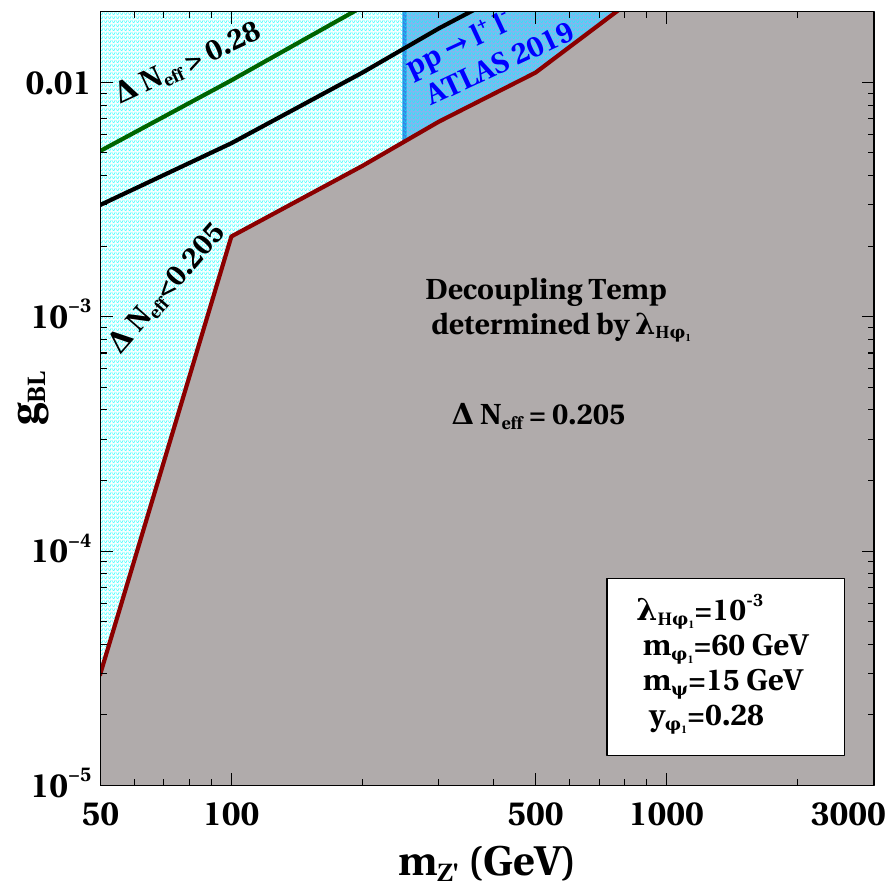}
\caption{ Bound on $m_{Z'}$ vs $g_{BL}$ for the WIMP scenario with different values of $\lambda_{H\phi_{1}}$. The solid black line separates regime 1 (below the solid black line) and regime 2 (above the solid black line). In all the plots, $m_{\phi_{1}}=60$ GeV, $m_{\psi}=15$ GeV, $y_{\phi_{1}}=0.28$.}
\label{fig:wimpbound}
\end{figure}


As we have discussed above, the kinetic decoupling of the dark sector ($\phi, \psi$ and $\nu_{R}$) from its bath is determined by both the Higgs-portal coupling and B-L gauge coupling. After the kinetic decoupling, due to large Yukawa coupling, dark sector maintain a dark equilibrium. An increase in $\lambda_{H\phi_{1}}$ coupling or $g_{BL}$ coupling keeps the dark sector in bath for longer. Increasing the $g_{BL}$ coupling to a very high value gives us a scenario where the dark matter freezes out before the decoupling of $\nu_{R}$. In this case, we do not have a separate dark sector evolution different from the SM bath. Hence, we have two regimes namely, regime 1: DM freezes out when $\xi_{\rm fo} \neq 1$ (i.e. separate dark sector); regime 2: DM freezes out when $\xi_{\rm fo} = 1$ (no separate dark sector).  Depending upon whether kinetic decoupling temperature is determined by gauged B-L coupling (sub-case 1) or Higgs-poral coupling (sub-case 2) we can have two further sub-cases.    

In the plots shown in Fig. \ref{fig:wimpbound}, we show the parameter space in $m_{Z'}$ versus $g_{BL}$ plane for WIMP type DM. Each of these plots correspond to a fixed value of $\lambda_{H\phi_{1}}$. The other parameters fixed in all the plots are, $m_{\phi_{1}}=60$ GeV, $m_{\psi}=15$ GeV, $y_{\phi_{1}}=0.28$. The solid black line separates regime 1 (upper left triangular region) from regime 2. The solid brown line separates the regime where kinetic decoupling temperature is determined by Higgs portal coupling from the one where it is determined by the gauge portal coupling. In the grey coloured region below the solid brown line, $g_{BL}$ and $m_{Z'}$ have no role in the kinetic decoupling of dark sector. In the upper panel plots, the region below the magenta line is where we have $\Delta N_{\rm eff} > 0.28$, disfavoured by Planck 2018 limits. In the grey coloured region, due to constant decoupling temperature, we have constant $\Delta N_{\rm eff}$. Due to smaller Higgs-portal coupling in the upper right panel plot compared to the upper left one, the grey coloured region shrinks. In the lower left panel plot, we increase the Higgs-portal coupling compared to upper left panel plot to $2.5 \times 10^{-4}$. For this particular choice of $\lambda_{H\phi_{1}}$, the region between the black and brown line has $\Delta N_{\rm eff} < 0.28$ and the region below the brown line has $\Delta N_{\rm eff}$=0.28, which is the maximum allowed value from Planck 2018 limits. In the lower right panel plot, $\lambda_{H\phi_{1}}$ is increased even further to $10^{-3}$ giving $\Delta N_{\rm eff} < 0.28$ in the whole regime 1. For regime 2, as DM freezes out before RHN decoupling, the bound from $\Delta N_{\rm eff}$ (denoted by region above the green line) is same as the bound shown in Fig. \ref{fig:thermal_Neff}. The entire parameter space shown in these plots satisfy the criteria of DM relic abundance. This is because DM abundance has very weak dependence on  the kinetic decoupling temperature. For the WIMP type DM scenario, we also check the constraints from direct detection and find them to be very weak due to radiative suppression. The details are given in appendix \ref{appen3}.

\section{Conclusion}
\label{sec:conclude}
We have studied a possible UV completion of the light Dirac neutrino portal dark matter scenario. In such a scenario, right chiral parts of light sub-eV Dirac neutrinos act like a portal between dark and visible sectors responsible for the production of dark matter. A gauged $U(1)_{B-L}$ symmetry provides one possible UV completion by naturally accommodating right chiral parts of neutrinos from anomaly cancellation requirements while also preventing direct coupling of DM, a gauge singlet Dirac fermion with the SM required for its stability. Keeping $U(1)_{B-L}$ symmetry breaking scale upto a few TeV ballpark, we study the details of dark matter production together with additional relativistic degrees of freedom $\Delta N_{\rm eff}$ brought in by right chiral parts of light Dirac neutrinos. While the chosen values of $U(1)_{B-L}$ gauge couplings ensures a non-zero thermal contribution to $\Delta N_{\rm eff}$, DM production can be either purely thermal or non-thermal depending upon the light Dirac neutrino portal Yukawa coupling. Although dark matter does not face stringent direct detection bounds due to loop-suppressed couplings with the SM quarks and charged leptons, the parameter space can be tightly constrained from other constraints related to structure formation, CMB constraints on $\Delta N_{\rm eff}$, collider constraints on $U(1)_{B-L}$ gauge bosons. We show interesting correlations in the parameter space from simultaneous requirement of correct DM phenomenology and $\Delta N_{\rm eff}$ indicating the region within reach of future CMB experiments.

\section*{Acknowledgements}
The work of DB is supported by the Science and Engineering Research Board (SERB), Government of India grant MTR/2022/000575. ND would like to thank Anirban Biswas, Dibyendu Nanda, Pritam Das, Sahabub Jahedi, and Suruj Jyoti Das for useful discussions. The work of ND is supported by the Ministry of Education, Government of India via the Prime Minister's Research Fellowship (PMRF) December 2021 scheme.

\appendix
{
\section{Anomaly cancellation in gauged $B-L$ model}
\label{appen_anomaly}
$U(1)_{B-L}$ gauge symmetry with only the SM fermions is not anomaly free. This is because the triangle anomalies for both $U(1)^3_{B-L}$ and the mixed $U(1)_{B-L}-(\text{gravity})^2$ diagrams are non vanishing. These triangle anomalies for the SM fermion content are given as
\begin{align}
\mathcal{A}_1 \left[ U(1)^3_{B-L} \right] = \mathcal{A}^{\text{SM}}_1 \left[ U(1)^3_{B-L} \right]=-3\,,  \nonumber \\
\mathcal{A}_2 \left[(\text{gravity})^2 \times U(1)_{B-L} \right] = \mathcal{A}^{\text{SM}}_2 \left[ (\text{gravity})^2 \times U(1)_{B-L} \right]=-3\,.
\end{align}
If three right handed neutrinos are added to the model, they contribute $\mathcal{A}^{\text{New}}_1 \left[ U(1)^3_{B-L} \right] = 3, \mathcal{A}^{\text{New}}_2 \left[ (\text{gravity})^2 \times U(1)_{B-L} \right] = 3$ leading to vanishing total of triangle anomalies. This is the most natural and economical $U(1)_{B-L}$ model studied extensively in the literature. However, there exist non-minimal ways of constructing anomaly free versions of $U(1)_{B-L}$ model. For example, it has been known for a few years that three right handed neutrinos with $B-L$ charges $5, -4, -4$ can also give rise to vanishing triangle anomalies \cite{Montero:2007cd} as follows.
\begin{align}
\mathcal{A}_1 \left[ U(1)^3_{B-L} \right] = \mathcal{A}^{\text{SM}}_1 \left[ U(1)^3_{B-L} \right]+\mathcal{A}^{\text{New}}_1 \left[ U(1)^3_{B-L} \right]=-3 + \left [ -5^3 - (-4)^3 - (-4)^3 \right]=0\,, \nonumber
\end{align}
\begin{align}
\mathcal{A}_2 \left[(\text{gravity})^2 \times U(1)_{B-L} \right] & = \mathcal{A}^{\text{SM}}_2 \left[ (\text{gravity})^2 \times U(1)_{B-L} \right]+ \mathcal{A}^{\text{New}}_2 \left[ (\text{gravity})^2 \times U(1)_{B-L} \right]\,, \nonumber \\
&=-3 + \left[ -5 - (-4) - (-4) \right]=0\,.
\end{align}
 Another solution to the anomaly cancellation conditions with irrational $B-L$ charges of new fermions was proposed by the authors of \cite{Wang:2015saa} where both DM and neutrino mass can have a common origin through radiative linear seesaw. Very recently, another anomaly free $U(1)_{B-L}$ framework was proposed where the additional right handed fermions possess more exotic $B-L$ charges namely, $-4/3, -1/3, -2/3, -2/3$ \cite{Patra:2016ofq}. These four chiral fermions constitute two Dirac fermion mass eigenstates, the lighter of which becomes the DM candidate having either thermal \cite{Patra:2016ofq} or non-thermal origins \cite{Biswas:2016iyh}. The light neutrino mass in this model arises from type II seesaw mechanism with the new chiral fermions playing no role in it. In  \cite{Nanda:2017bmi}, such chiral fermions with fractional charges were also responsible for generating light neutrino masses at one loop level. In the recent work on $U(1)_{B-L}$ gauge symmetry with two component DM \cite{Bernal:2018aon}, the authors considered two right handed neutrinos with $B-L$ number -1 each so that the model still remains anomalous. The remaining anomalies were cancelled by four chiral fermions with fractional $B-L$ charges leading to two Dirac fermion mass eigenstates both of which are stable and hence DM candidates. In \cite{Biswas:2019ygr}, type III seesaw was implemented in in $U(1)_{B-L}$ model while introducing additional chiral fermions to keep the model anomaly free. Appropriate choice of chiral fermions also leads to multi-component DM in such $U(1)_{B-L}$ symmetric type III seesaw scenario.}

 {While it is possible to study Dirac neutrino portal DM in gauged $U(1)_{B-L}$ frameworks with different anomaly free combinations of chiral fermions, it will require additional scalar fields in order to be consistent with the desired DM phenomenology and non-zero Dirac neutrino mass. For example, if we consider three right handed neutrinos with $B-L$ charges $5, -4, -4$, we will require two additional Higgs doublets of $U(1)_{B-L}$ charge $6, 3$ respectively in order to generate Dirac neutrino masses at tree level. Two singlet scalars $\phi_{1,2}$ and DM $\psi$, similar to our model, can be incorporated to generate the Dirac neutrino portal of DM and break $U(1)_{B-L}$ symmetry spontaneously. Similar non-minimal scalar content will be required for other chiral fermions mentioned above. While the calculations will be more involved due to more particles, the generic conclusions reached in our work should not change significantly.}

\section{Calculation of FSL}\label{appen_FSL}

The free-streaming length (FSL) can be quantified as \cite{Schneider:2011yu}-
\begin{eqnarray} \label{free-streaming}
    \lambda_{\rm FSL} = \bigintss_{t_{\rm prod}}^{t_{\rm eq}}
\dfrac{\langle v\rangle}{a} \,dt = \bigintss_{T_{\rm prod}}^{T_{\rm eq}}
\dfrac{\langle v(T)\rangle}{a(T)} \dfrac{dt}{dT}\,dT.
\end{eqnarray}
Here $t_{\rm prod}$ ($T_{\rm prod}$) is the time (temperature) when maximum production of dark matter occurs. $t_{\rm eq}$ ($T_{\rm eq}$) is the time (temperature) of matter-radiation equality after which the structure formation starts. In terms of distribution function $f_{\psi}(q_{1},T)$, the average velocity $\langle v(T)\rangle$ can be written as 
\begin{eqnarray}
\langle{v(T)}\rangle =
\dfrac{\bigintss \frac{q_{1}}{E_1}\,
\frac{d^3 q_1}{(2\pi)^3}\,f_{\psi}(q_1,T)}
{\bigintss \frac{d^3 q_1}{(2\pi)^3}
\,f_{\psi}(q_1,T)}.
\end{eqnarray}
Here, $q_{1}$ and $E_{1}$ denotes the momentum and energy of dark matter respectively. To calculate the distribution function $f_{\psi}(q_{1},T)$, it is convenient to change these variables to $r$ and $\xi_{\psi}$ given by
\begin{eqnarray}
    r &=& \frac{m_{0}}{T}, \\ \nonumber
    \xi_{\psi} &=& \left(\frac{g^{s}_{*}(T_{0})}{g^{s}_{*}(T)}\right)^{1/3} \frac{q_{1}}{T},    
\end{eqnarray}
where $m_{0}$ and $T_{0}$ is some reference mass and temperature respectively. For calculating $f_{\psi}(q_{1},T)$, one also needs to calculate the distribution function of $\phi_{1}$, $f_{\phi_{1}}(p,T)$.

For the process $\phi_{1}(p_{1}) \to \psi(q_{1}) + \nu_{R}(q_{2})$, the Boltzmann equation for the distribution function of $\psi$ can be written as
\begin{align}
\frac{\partial f_{\psi}}{\partial t} - \mathcal{H} q_1 \frac{\partial f_{\psi}}{\partial q_1} =& \frac{1}{16\pi\,E_{q_1}\,q_1}
\int_{p^{\rm min}_1}^{p^{\rm max}_1} \dfrac{p_1 dp_1}{E_{p_1}}
\rvert \mathcal{M} \rvert^2_{\phi \to \bar{\nu}_R \psi}
\,f_{\phi_{1}}(p_1).
\end{align}
The expressions for $p_{1}^{\rm max}$ and $p_{1}^{\rm min}$ for $m_{\phi_{1}} \gg m_{\psi}$ are
\begin{eqnarray}
p^{\rm min}_1 &\simeq& \dfrac{m^2_{\phi_{1}}}{2\,m^2_{\psi}}
\left(-q_1 +\sqrt{q_1^2 - 4\frac{m^2_{\psi}}{m_{\phi_{1}}^2} q_1^2 + m^2_{\psi}}\right)\,,\\
p^{\rm max}_1 &\simeq& \dfrac{m^2_{\phi_{1}}}{2\,m^2_{\psi}}
\left(q_1  + \sqrt{q_1^2  - 4\frac{m^2_{\psi}}{m_{\phi_{1}}^2} q_1^2 + m^2_{\psi}}\right)\,.
\end{eqnarray}

The distribution function for $\phi_{1}$ after its freeze-out can be obtained from 

\begin{align}
    \frac{\partial f_{\phi_{1}}}{\partial t} - \mathcal{H} p_1 \frac{\partial f_{\phi_{1}}}{\partial p_1} = -\frac{1}{2E_{p_1}} &\int \frac{d^3q_1}{2E_{q_1}(2\pi)^3}\frac{d^3q_2}{2E_{q_2}(2\pi)^3} (2\pi)^4 \delta^4(P_1-Q_1-Q_2) \nonumber \\
    &\times  \lvert \mathcal{M} \rvert^2_{\phi_{1}\to \psi \Bar{\nu}_{R}} f_{\phi_{1}}(p_{1})\, \\ \nonumber
    =- f_{\phi_{1}} & \frac{m_{\phi_{1}}}{\sqrt{p_{1}^2 + m_{\phi_{1}}^2}} \Gamma_{\phi_{1} \to \psi \Bar{\nu}_{R}}.
\end{align}Here $P_{1}, Q_{1}$ and $Q_{2}$ are the four momenta corresponding to $\phi_{1}$, $\psi$ and $\nu_{R}$ respectively. With these above equations and using the transformation $ r = \frac{m_{0}}{T}, \xi_{\psi} = \left(\frac{g^{s}_{*}(T_{0})}{g^{s}_{*}(T)}\right)^{1/3} \frac{q_{1}}{T}$, we obtain the distribution function $f_{\psi}(\xi_{\psi}, r)$ \cite{Konig:2016dzg, Biswas:2016iyh, Biswas:2022vkq}. With these, the average velocity of DM and free-streaming length can be re-expressed in terms of new variables as
\begin{eqnarray}
\langle v(r)\rangle = \dfrac{\mathcal{A}(r)}
{\bigintss \xi^2_{\psi}\,f_{\psi}(\xi_{\psi},r)\,d\xi_{\psi}}\times
{\bigintss \dfrac{\xi^3_{\psi}\,f_{\psi}(\xi_{\psi},r)\,
d\xi_{\psi}}{\sqrt{(\mathcal{A}(r) \xi_{\psi})^2 
+ (\frac{r}{m_0}\,m_{\psi})^2}}}\, \\
\lambda_{\rm FSL} = 
\left(\dfrac{11}{43}\right)^{1/3}\,r_0
\int_{r_{\rm prod}}^{r_{\rm eq}}
{\langle{v(r)}\rangle\,g_s^{1/3}}
\frac{\beta}{\mathcal{H}(r)} \frac{dr}{r^2}, 
\end{eqnarray}
where $ \mathcal{A}(r) = \left(\frac{g_{*}^{s}(m_0/r)}{g_{*}^{s}(m_0/T_0)}\right)^{1/3}$.

\section{{Calculation of LHC bound on $g_{BL} -  m_{Z'}$ plane}} \label{appen_gBL_mz'}

We evaluate the constraint on $U(1)_{B-L}$ gauge coupling $g_{BL}$ from dilepton channel: $pp \to Z' \to l^+ l^-$. Using the narrow-width approximation, the cross-section $pp \to Z'$ can be written as  
    \begin{equation}
        \sigma(pp \to Z') = 2 \sum_{q,\Bar{q}} \int dx \int dy f_{q}(x,Q) f_{\Bar{q}}(x,Q) \hat{\sigma}(\hat{s}), 
    \end{equation}
    where $\hat{s}=xys$ with $f_{q}$, $f_{\Bar{q}}$ being the parton distribution functions for quarks and anti-quark respectively. The narrow-width approximation cross section $\hat{\sigma}(\hat{s})$ is given as 
    \begin{equation}
        \hat{\sigma}(\hat{s}) =  \frac{4 \pi^2}{3} \frac{\Gamma(Z' \to q\Bar{q})}{m_{Z'}} \delta(\hat{s}-m_{Z'}^2).
    \end{equation}
    We use the package \texttt{CalcHEP} for the computation of $\sigma(pp\to Z')$ where CTEQ6L is used for parton distribution functions. 

    In our model, apart from quarks and leptons, three $\nu_{R}$s and two scalars $\phi_{1}$ and $\phi_{2}$ are charged under $U(1)_{B-L}$. The decay of $Z'$ to both the BSM scalars $\phi_{1}$ and $\phi_{2}$ is not possible due to kinematic restrictions imposed. The decay width of $Z'$ to quarks, charged leptons, SM neutrinos and $\nu_{R}$ can be written as
    \begin{eqnarray}
        \Gamma(Z' \to q\Bar{q}) \simeq 12\times3\times\frac{1}{9} \frac{g^2_{BL}}{24\pi} m_{Z'} = 4 \times \frac{g^2_{BL}}{24\pi} m_{Z'} \\ \nonumber
        \Gamma(Z' \to l^{-}l^{+}) \simeq 6 \times  \frac{g^2_{BL}}{24\pi} m_{Z'} \\ \nonumber
        \Gamma(Z' \to \nu_{L}\Bar{\nu}_{L}) \simeq 3 \times  \frac{g^2_{BL}}{24\pi} m_{Z'} \\ \nonumber
        \Gamma(Z' \to \nu_{R}\Bar{\nu}_{R}) \simeq 3 \times  \frac{g^2_{BL}}{24\pi} m_{Z'}
        \end{eqnarray}
    respectively, assuming massless final states. With this, we can calculate the cross-section $\sigma (pp \to Z' \to l^+ l^-) \approx \sigma(pp\to Z')\, {\rm BR}(Z' \to l^{-}l^{+})$ for particular values of $m_{Z'}$ and $g_{BL}$. As $\sigma(pp\to Z')$ is proportional to $g^2_{BL}$, we can normalise it by dividing with $g^2_{BL}$.
\begin{figure}
        \includegraphics[height=5cm, width=8cm]{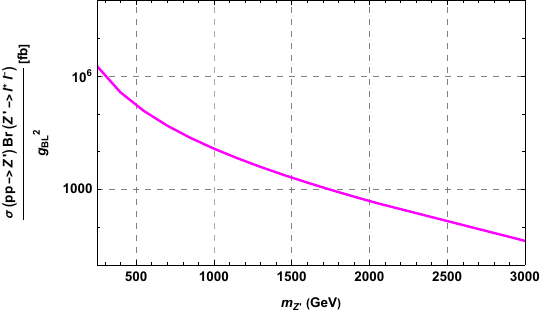}
        \includegraphics[height=5cm, width=8cm]{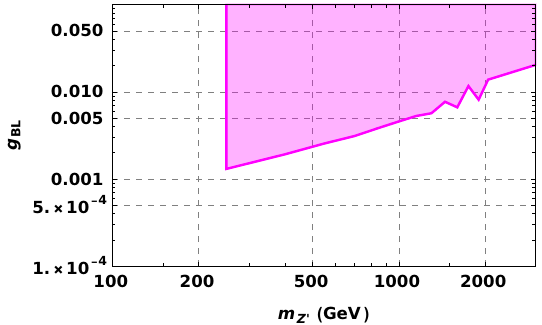}
        \caption{Left panel: $\frac{\sigma(pp\to Z') \, {\rm BR}(Z' \to l^{-}l^{+})}{g^2_{BL}}$ vs $m_{Z'}$. Right panel: Bound on $g_{BL}$ vs $m_{Z'}$ plane.}
        \label{fig:bound_gbl}
    \end{figure}
    
    The left panel plot of Fig. \ref{fig:bound_gbl} shows $\frac{\sigma(pp\to Z') \, {\rm BR}(Z' \to l^{-}l^{+})}{g^2_{BL}}$ as a function of $m_{Z'}$ for our model.
    In the right panel plot, we show the bound on $U(1)_{B-L}$ gauge coupling for our model in magenta region. The bound is obtained by comparing the left plot with the final results from ATLAS collaboration of the LHC Run-2 with $139$  $\text{fb}^{-1}$ integrated luminosity. 
\color{black}
\section{Estimation of dark sector temperature}\label{appen_dark_temp}

Here we provide an estimation of dark sector temperature using the entropy conservation. Let us assume that $g^{\rm s,DS}_{*}$ and $g^{\rm s}_{*}$ are the effective number of relativistic degree of freedom of dark sector and standard model respectively. At the epoch of decoupling of dark sector from the standard model bath, the scale factor is taken as $a_{\rm dec}$. Using the conservation of entropy, we get
\begin{eqnarray}
    g^{\rm s}_{\rm *,dec} a^3_{\rm dec} (T_{\rm dec})^3 = g^{\rm s}_{*,\rm aft} a^3_{\rm aft}(T_{\rm aft})^3 \nonumber \\
    g^{\rm s,DS}_{\rm *,dec} a^3_{\rm dec} (T^{\rm DS}_{\rm dec})^3 = g^{\rm s,DS}_{*,\rm aft} a^3_{\rm aft}(T^{\rm DS}_{\rm aft})^3,
\end{eqnarray}
where $a_{\rm aft}$ denotes any period after decoupling and $T^{\rm DS}$ and $T$ denote temperature of dark sector and SM bath respectively. At the epoch of decoupling, both the dark sector and SM bath have same temperature, $T_{\rm dec}=T^{\rm DS}_{\rm dec}$. Hence, from the above equations, we get -
\begin{eqnarray}
    \left(\frac{T^{\rm DS}_{\rm aft}}{T_{\rm aft}}\right)^4 = \left(\frac{g^{\rm s,DS}_{\rm *,dec}}{g^{\rm s,DS}_{\rm *,\rm aft}}\frac{g^{\rm s}_{\rm *, \rm aft}}{g^{\rm s}_{\rm *,dec}}\right)^{4/3}.
\end{eqnarray}

For the situation where dark sector consists of $\phi_{1}, \psi$ and $\nu_{R}$, we have $g^{\rm s,DS}_{*,\rm dec} = 2 + 6\times \frac{7}{8} + 4\times\frac{7}{8} = 10.75$. Taking $a_{\rm aft}$ to a period when both $\phi_{1}$ and $\psi$ annihilate or become non-relativistic, we get $g^{\rm s,DS}_{*, \rm aft} = 6\times \frac{7}{8} = 5.25$. If during this period, $g^{\rm s}_{*}$ remains constant, we get
\begin{eqnarray}
    \left(\frac{T^{\rm DS}_{\rm aft}}{T_{\rm aft}}\right)^4 = \xi^4 = \left(\frac{10.75}{5.25}\right)^{4/3} = 2.6.
\end{eqnarray}

So, at maximum, we can expect a 2.6 times increase in dark sector temperature from the SM bath. The numerical results shown in Fig. \ref{fig:LY_neff} are consistent with this analytical estimation.

However, we can have a value of $g^{\rm s,DS}_{\rm *,dec}<10.75$ at the epoch of decoupling. This is because before decoupling, some fraction of dark sector particles can become non-relativistic. For example, let us consider $m_{\phi_{1}}=60$ GeV, $m_{\psi}=15$ GeV. Decoupling above $T \sim 60$ GeV gives $g^{s,\rm DS}_{*,\rm dec}\sim 10.75$ whereas decoupling after $T \sim 60$ GeV (e.g. $10$ GeV) gives $g^{\rm s,DS}_{*,\rm dec}<10.75$. This results in a smaller increase in dark sector temperature from SM bath. The same behaviour can also been seen from Fig. \ref{fig:LY_neff}. 

\section{Direct Detection of WIMP type DM}
\label{appen3}
\begin{figure}[h!]
\includegraphics[height=8 cm, width =9 cm]{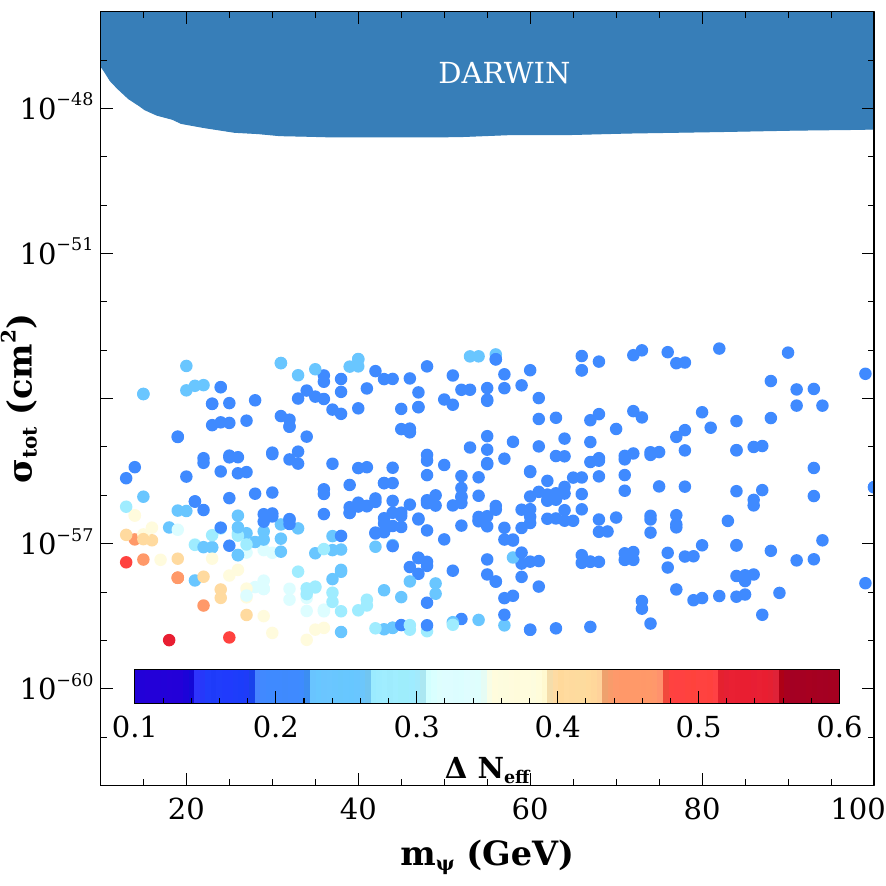}
\caption{The DM-nucleon cross-section vs DM mass as a function of effective number of relativistic species. The blue color region shows the sensitivity of DARWIN experiment \cite{DARWIN:2016hyl}.}
\label{fig:DD}
\end{figure}

In this setup, dark matter $\psi$ does not interact with the nucleus via tree level diagram. However, via one loop diagram, it can scatter off the nucleus. We get two diagrams, one where the mediator is SM Higgs and the other where the mediator is B-L gauged boson. For the Higgs mediated diagram, the relevant interaction vertices are $g_{\psi\Bar{\psi}h} \psi\Bar{\psi}h$ and $\frac{m_{q}}{v}q\Bar{q}$. Here $q$ represents the SM quarks. The effective interaction term for Higgs mediated diagram can be written as 
\begin{eqnarray}
    \mathcal{L}^{h}_{eff} = \frac{m_{q}}{v}\frac{1}{m_{h}^2}g_{\psi\Bar{\psi}h}\psi\Bar{\psi}q\Bar{q}.
\end{eqnarray}
Similarly for the $Z'$ mediated diagram, the effective interaction term can be written as 
\begin{eqnarray}
    \mathcal{L}^{Z'}_{eff} = -\frac{g_{BL}}{3}\frac{1}{m_{Z'}^2}g_{\psi\Bar{\psi}Z'}\psi\gamma^{\mu}\Bar{\psi}q \gamma_{\mu}\Bar{q}.
\end{eqnarray}

The interaction vertices $g_{\psi\Bar{\psi}h}$ and $g_{\psi\Bar{\psi}Z'}$ for one loop diagram are calculated using Package-X\cite{Patel:2016fam} and their expressions are given as follows (for zero momentum transfer)\footnote{We consider only one of the two different one-loop diagrams to calculate the effective $g_{\psi\Bar{\psi}Z'}$ vertex for simplicity.}
\begin{eqnarray}
    g_{\psi\Bar{\psi}h} &=& \frac{i}{16 \pi^2} y^{2}_{\phi_{1}} \lambda_{H\phi_{1}} v \frac{1}{m_{\psi}}\left[1+\left(\frac{m^2_{\phi_{1}}}{m^2_{\psi}}-1 \right) \text{ln} \left(1-\frac{m^2_{\psi}}{m^2_{\phi_{1}}}\right)\right] \nonumber \\
    g_{\psi\Bar{\psi}Z'} &=& \frac{i}{16 \pi^2} y^{2}_{\phi_{1}} g_{BL} \left[\frac{3}{2} \frac{m^2_{\phi_{1}}}{m^2_{\psi}}  + \left(\frac{m^2_{\phi_{1}}}{m^2_{\psi}}-1 \right) \text{ln} \left(1-\frac{m^2_{\psi}}{m^2_{\phi_{1}}}\right)  \left(\frac{3}{2} \frac{m^2_{\phi_{1}}}{m^2_{\psi}}  
 +   \frac{1}{2}\right)\right].
\end{eqnarray}

The total DM-nucleon cross-section can be written as \cite{Biswas:2016ewm, Bhattacharya:2022vxm, Jungman:1995df} - 
\begin{eqnarray}
    \sigma_{\rm tot} = \frac{1}{\pi} \frac{m_{N}^2 m_{\psi}^2}{(m_{N}+m_{\psi})^2} \left[\frac{m_{N}}{v}\frac{1}{m_{h}^2} g_{\psi\Bar{\psi}h} f_{N} + \frac{g_{BL}}{3}\frac{1}{m_{Z'}^2}g_{\psi\Bar{\psi}Z'}f_{Z'}\right]^2,
\end{eqnarray}
where $f_{N} \sim 0.3$, $f_{Z'}=3$ and $m_{N}$ is the nucleon mass. In Fig. \ref{fig:DD}, we show the DM-nucleon cross-section as a function of $m_{\psi}$. To calculate the cross-section, we take the values of different parameters same as for the resultant points we obtained in Fig. \ref{fig:LY_scan}. All the points satisfy correct DM relic abundance and range of different parameters for these points are given by Eq. \eqref{WIMP_parameters}. We find that the total cross-section is way below the direct detection limit.

\bibliographystyle{JHEP}
\bibliography{ref.bib, ref1.bib} 

\end{document}